\documentclass[12pt]{iopart}
\usepackage{graphicx}
\usepackage{todonotes}
\usepackage{pstricks}
\usepackage{pstricks-add}
\usepackage{soul}
\usepackage{xcolor}

\begin{document}

\title{Effect of interactions and non-uniform magnetic states on the magnetization reversal of iron nanowire arrays}

\author{I.~S.~Dubitskiy}
\address{Petersburg Nuclear Physics Institute named by B.P. Konstantinov of National Research Centre "Kurchatov Institute", 1, Orlova roscha mcr.,  188300, Gatchina, Leningrad Region, Russia}

\author{A.~H.~A.~Elmekawy}
\address{Saint Petersburg State University, 198504 St. Petersburg, Russia}

\author{E.~G.~Iashina}
\address{Petersburg Nuclear Physics Institute named by B.P. Konstantinov of National Research Centre "Kurchatov Institute", 1, Orlova roscha mcr.,  188300, Gatchina, Leningrad Region, Russia}

\author{S.~V.~Sotnichuk}
\address{Saint Petersburg State University, 198504 St. Petersburg, Russia}
\address{Lomonosov Moscow State University, 119991 Moscow, Russia}

\author{K.~S.~Napolskii}
\address{Lomonosov Moscow State University, 119991 Moscow, Russia}

\author{D.~Menzel}
\address{Institute of Condensed Matter Physics, 38106 Braunschweig, Germany}

\author{A.~A.~Mistonov}
\address{Saint Petersburg State University, 198504 St. Petersburg, Russia}

\vspace{10pt}
\begin{indented}
\item[]April 2020
\end{indented}

\begin{abstract}
Ordered ferromagnetic nanowire arrays are widely studied due to the diversity of possible applications. However, there is still no complete understanding of the relation between the array's parameters and its magnetic behavior. The effect of vortex states on the magnetization reversal of large-diameter nanowires is of particular interest. Here, we compare analytical and micromagnetic models with experimental results for three arrays of iron nanowires with diameters of 33, 52 and 70 nm in order to find the balance between the number of approximations and resources used for the calculations. The influence of the vortex states and the effect of interwire interactions on the remagnetization curves is discussed. It has been found that 7 nanowires treated by a mean field model are able to reproduce well the reversal behaviour of the whole array in the case of large diameter nanowires. Vortex states tend to decrease the influence of the structural inhomogeneities on reversal process and thus lead to the increased predictability of the system.    

 \end{abstract}

\section{Introduction}
\label{sec:introduction}

The list of possible applications of magnetic nanowires is continuously expanding~\cite{DEMIREL2015104, supercond, SU20091062, Wierzbicki_2015, parkin2008magnetic}, that requires a deeper understanding of the remagnetization mechanism at the nanoscale. Long-term and fruitful studies of nanowire arrays have significantly improved the understanding of their magnetic properties~\cite{bruck2018handbook, vazquez2015magnetic, zeng2002structure, ortega2017magnetic}. The first models describing the magnetic properties of ferromagnetic nanowire arrays have been based on simple but convenient assumptions. The model of coherent rotation has been used to describe the remagnetization process and the interactions in the array have been taken into account by considering the nanowires as point dipoles~\cite{ishii1989magnetic, sun2005tuning}. However, the improvement of synthesis methods~\cite{goncharova2017oriented, schlesinger2011modern, ivanov2013magnetic, schaefer2016nico, pitzschel2011magnetic, ruiz2019tailoring} as well as the new capabilities to investigate individual wires~\cite{bran2018magnetization, wolf2019holographic, biziere2013imaging} have shown that these models are often oversimplified. It has been found that the magnetization reversal process in low-anisotropy soft nanowires usually occurs via the motion of the domain wall if the field is applied along the nanowire axis~\cite{bruck2018handbook}. The type of the domain wall is determined by nonuniform states that are present in the nanowire which in turn are connected with nanowire diameter ($D$) to exchange length ($l_{exch}$) ratio. The exchange length is defined as $l_{exch}=\sqrt{2A/(\mu_0 M^2_S)}$, where $\mu_0$ is a magnetic constant, $A$ is an exchange stiffness constant and $M_S$ is saturation magnetization. If the wire diameter is less than $7l_{exch}$, remagnetization should occur due to movement of the transverse domain wall (TDW)~\cite{forster2002domain, wieser2004domain}. Otherwise, the vortex (Bloch point) domain wall (VDW) may arise~\cite{thiaville2006domain}. Pinning of the domain walls in real systems may require some refinement of the models~\cite{ebels2000spin, da2016nucleation, stavno2017probing}. A more detailed picture can be in principle obtained by micromagnetic simulations which can be applied to systems that can not be handled by analytic models. But some issues arise along the way in this case as well. For instance Bloch-point that is always present in VDW can not be tackled in continuous models. Applying numerical simulations to systems with VDW can lead to some undesirable artifacts~\cite{thiaville2003micromagnetic}. Micromagnetic simulations are also limited by the number of nanowires that can be considered. This may decrease the accuracy of the results because of long range nature of inter wires interaction. However, in recent years, some progress has been made in this area~\cite{fangohr2009new}. Nevertheless latterly exciting analytical models of nanowire arrays remagnetization have been proposed~\cite{vock2017role, da2011reduction, wang2008magnetic, bochmann2018preparation, vivas2012magnetic, lavin2009angular}. It is interesting to compare them with micromagnetic simulations and experiments in the case of some new systems. The purpose of this article is to match all these models and find an approach that would maintain a balance between computational costs, simplicity and agreement with the experiment. Deeper understanding of remagnetization behavior of nanowire arrays may be relevant for the design of new nanomagnetic systems and the needs of spintronics~\cite{bran2018magnetization} or medical applications~\cite{egolf2016hyperthermia, leulmi2015triggering}.

In order to compare several approaches, we have chosen iron nanowires as a testing system. Iron is a suitable material due to several reasons. It possesses a small exchange length that leads to the arising of non-uniform states even in the case of nanowires with small diameters~\cite{bonilla2017magnetic}. Moreover large saturation magnetization should enhance inter wire interactions. 
	
We consider the case when the nanowire diameter substantially exceeds the iron exchange length of 3.5~nm although nanowires still remain nanoscale. Three different $D/l_{exch}$ ratios of 10 (D~=~33~nm), 15 (D~=~52~nm) and 20 (D~=~70~nm) have been reviewed. The period of the structure $a$ is 101 nm, which corresponds to the most common interpore distance of porous anodic alumina used as a template for the preparation of iron nanowires. The porosity value $p=\pi D^2/ (2\sqrt{3} a^2)$ in this case varies from 0.097 to 0.44.

The branches of the hysteresis loops for the studied arrays of nanowire are very close to straight lines. In this regard, it is convenient to compare the main characteristics of these loops, namely: saturation field, coercivity and so called interaction field. The latter can be defined as the difference between saturation field and coercivity~\cite{bochmann2018preparation}.

\section{Samples prepartion}
\label{sec:samples}

\begin{figure}[hbtp]
	\begin{minipage}{0.7\linewidth}
		\center{\includegraphics[width=1\linewidth]{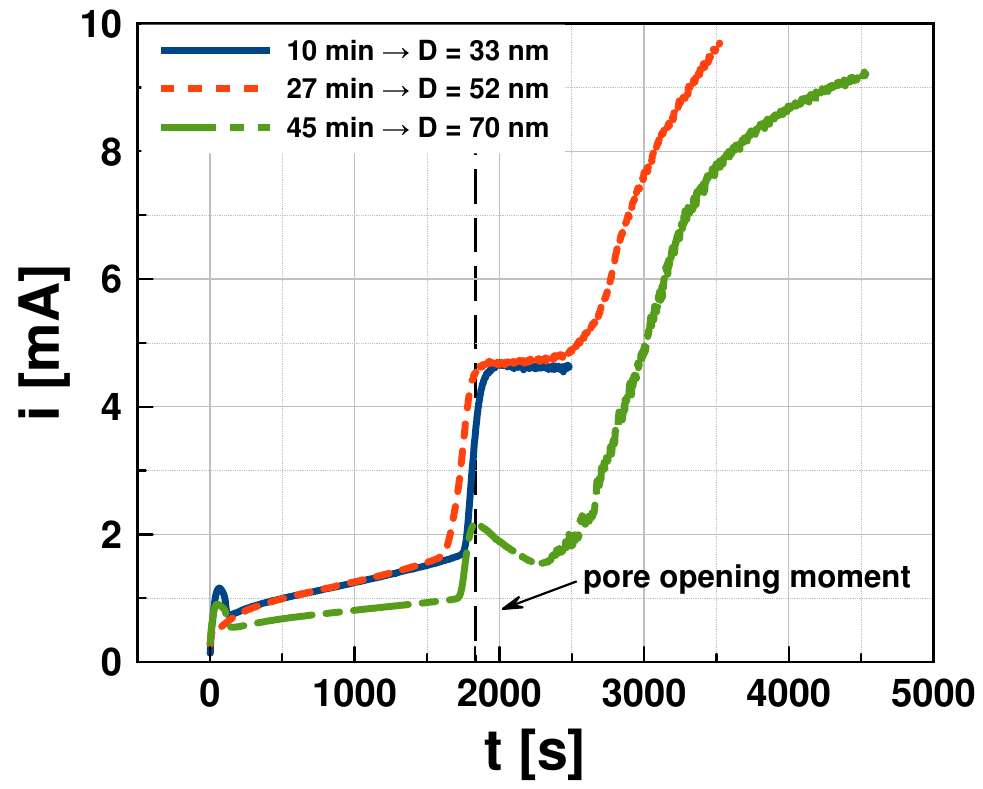}}
	\end{minipage}\\
	\caption{Current transients registered during barrier layer etching.}
	\label{ris:samples_prep}
\end{figure}

In order to prepare ordered arrays of iron nanowires, a templated electrodeposition technique with the use of porous anodic aluminum oxide (AAO) as a template was used. At first, 100-$\mu$m-thick aluminum foil (99.99\%) was electropolished in the solution containing 1.85 M CrO$_3$ and 13 M H$_3$PO$_4$, at 80$^{\circ}$~C for providing a smooth metal surface. Then aluminum was anodized using a two-step anodization technique in 0.3 M H$_2$C$_2$O$_4$ at 40 V. For this purpose, a two-electrode electrochemical cell with the Pt wire as a counter electrode was used. The temperature of the electrolyte was maintained constant at 0-3$^{\circ}$~C. After the first anodization step, a sacrificial AAO layer with a thickness of 10~$\mu$m was etched away in an aqueous solution of CrO$_3$ and H$_3$PO$_4$. Then the second anodization was performed to form AAO porous film with a thickness of 35~$\mu$m. Aluminum remained after the anodization was dissolved in the solution of Br$_2$ in CH$_3$OH (1:10 vol.). Finally, a barrier oxide layer was etched in 3 M H$_3$PO$_4$ using the electrochemical detection technique of pore opening moment~\cite{lillo2009pore}. This method allows one to control the pores final diameter by holding the membrane in the acid for a certain time after the pore opening moment (Fig.~\ref{ris:samples_prep}). The etching times after the pore opening moment of 10, 27, and 45 minutes correspond to the AAO templates with the pore diameters of 33, 52, and 70 nm, respectively. At the last stage of template fabrication, a 200-nm-thick Au layer was deposited at the bottom side of the AAO templates by magnetron sputtering.

Iron was electrodeposited from an electrolyte containing 0.5 M FeSO$_4$, 0.5 M Na$_2$SO$_4$, 0.4 M H$_3$BO$_3$, and 0.006 M ascorbic acid at room temperature in a three-electrode electrochemical cell. AAO with sputtered Au served as a working electrode, a Pt wire ring was used as a counter electrode, a saturated (KCl) Ag/AgCl electrode connected with the cell via Luggin-Haber capillary was a reference electrode. A deposition potential of –0.8 V, as well as a short (0.1 s) nucleation potential pulse of –1.2 V, was applied using Autolab PGSTAT101 potentiostat. During the electrodeposition, the electrolyte was rigorously agitated.

\section{Methods}
\label{sec:methods}

\subsection{Samples characterization}
\label{subsec:characterization}

\begin{figure*}[hbtp]
	\begin{minipage}{0.5\linewidth}
		\center{\includegraphics[width=1\linewidth]{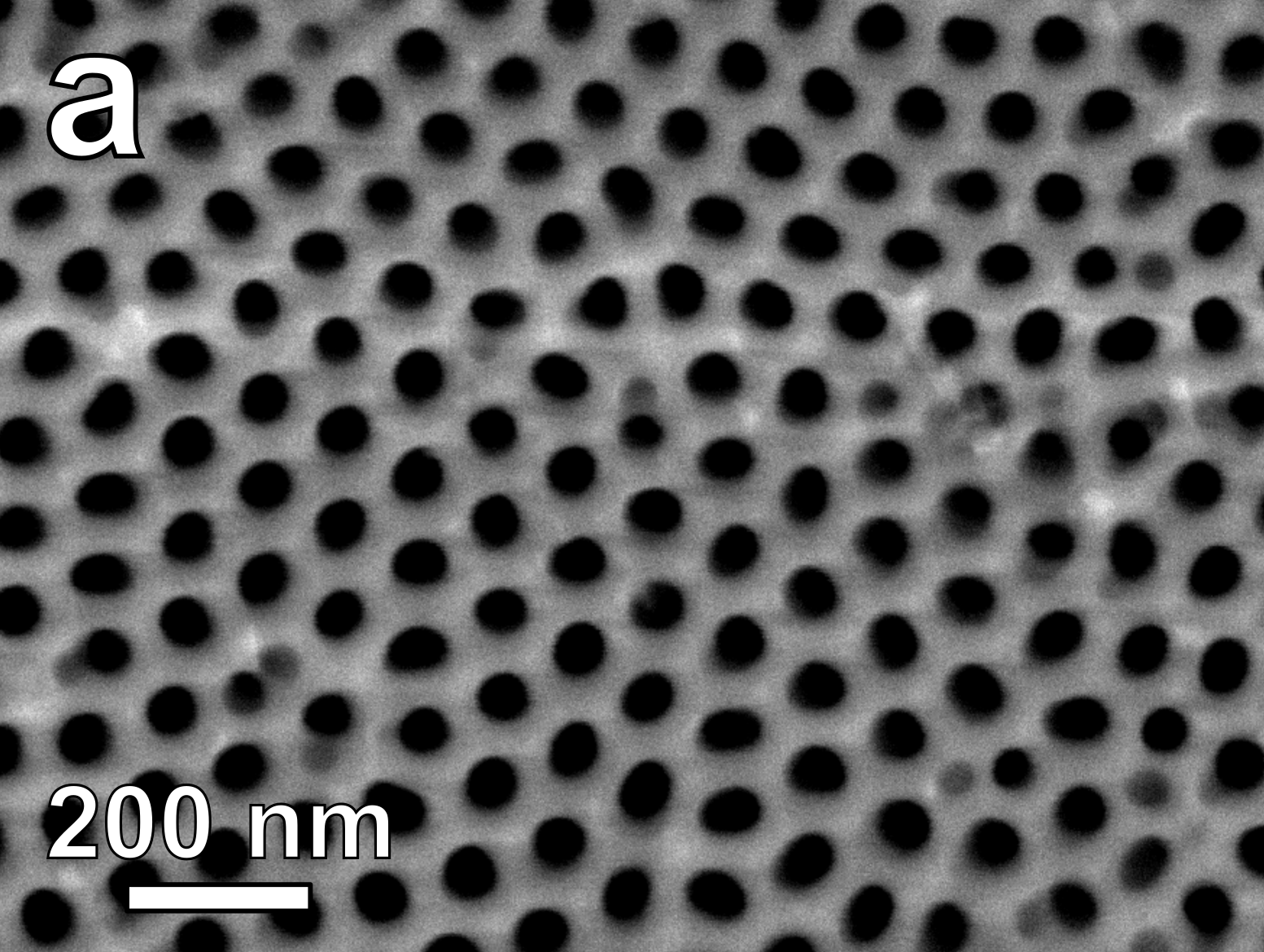}}
	\end{minipage}
    \begin{minipage}{0.5\linewidth}
		\center{\includegraphics[width=1\linewidth]{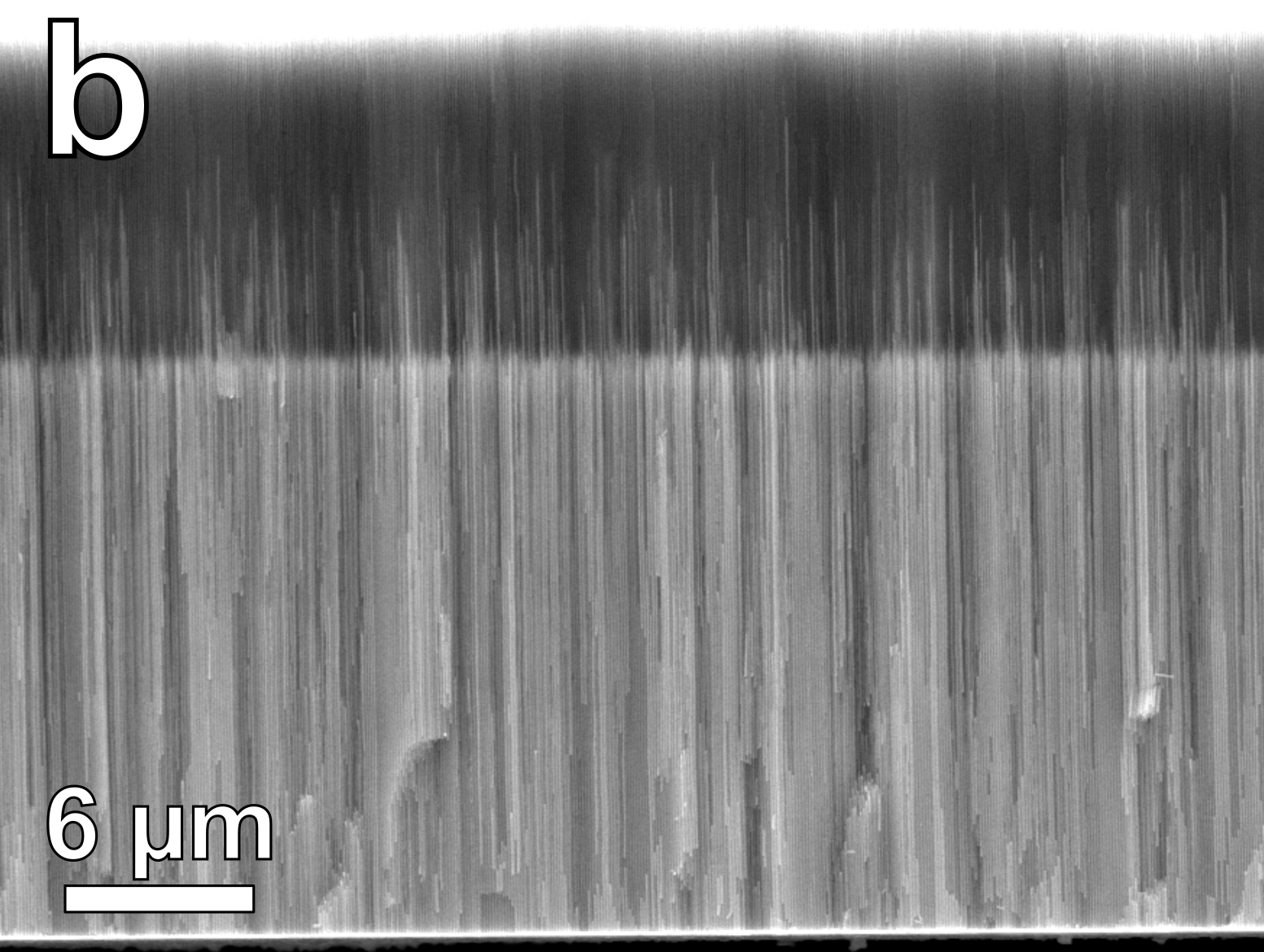}}
	\end{minipage}
	\caption{SEM images of the top side of the AAO template with a pore diameter of 52 nm (a) and the cross-section of corresponding Fe/AAO nanocomposite (b).}
	\label{ris:sem}
\end{figure*}

The morphology of iron nanowire arrays was studied by scanning electron microscopy (SEM) using a Zeiss AURIGA Laser microscope. Fig.~\ref{ris:sem} demonstrates the top side of the AAO template and a cross-section of Fe/AAO nanocomposite. The AAO template possesses a well-ordered structure with uniform pores. To determine the pore diameters and the distances between their centers, image processing with implementing the Voronoi algorithm was carried out in Statistics2D software~\cite{roslyakov2017growth}. Experimental distributions were approximated by the Gauss function. All geometrical parameters of Fe/AAO nanocomposites are presented in Table~\ref{tab:sem_data} (the samples are denoted as Fe$_x$, where $x$ is the estimated pore diameter in nanometers).

\begin{table}[hbtp]
    \centering
    \resizebox{0.5\columnwidth}{!}{
    \begin{tabular}{c|c|c|c}
        Sample & D,~nm & a,~nm & L,~$\mu$m \\
        \hline
       Fe$_{33}$  & 33~$\pm$~2 &  & 30.1~$\pm$~0.7\\
       Fe$_{52}$ & 55~$\pm$~4 & 101~$\pm$4 & 18.8~$\pm$~0.6\\
       Fe$_{70}$ & 69~$\pm$~7&  & 26.9~$\pm$~0.7
    \end{tabular}
    }
    \caption{Diameter (D), distance between centers of nanowires (a), and average nanowire length (L) for the Fe/AAO nanocomposites according to SEM data.}
    \label{tab:sem_data}
\end{table}

\begin{figure}[hbtp]
	\begin{minipage}{0.5\linewidth}
		\center{\includegraphics[width=1\linewidth]{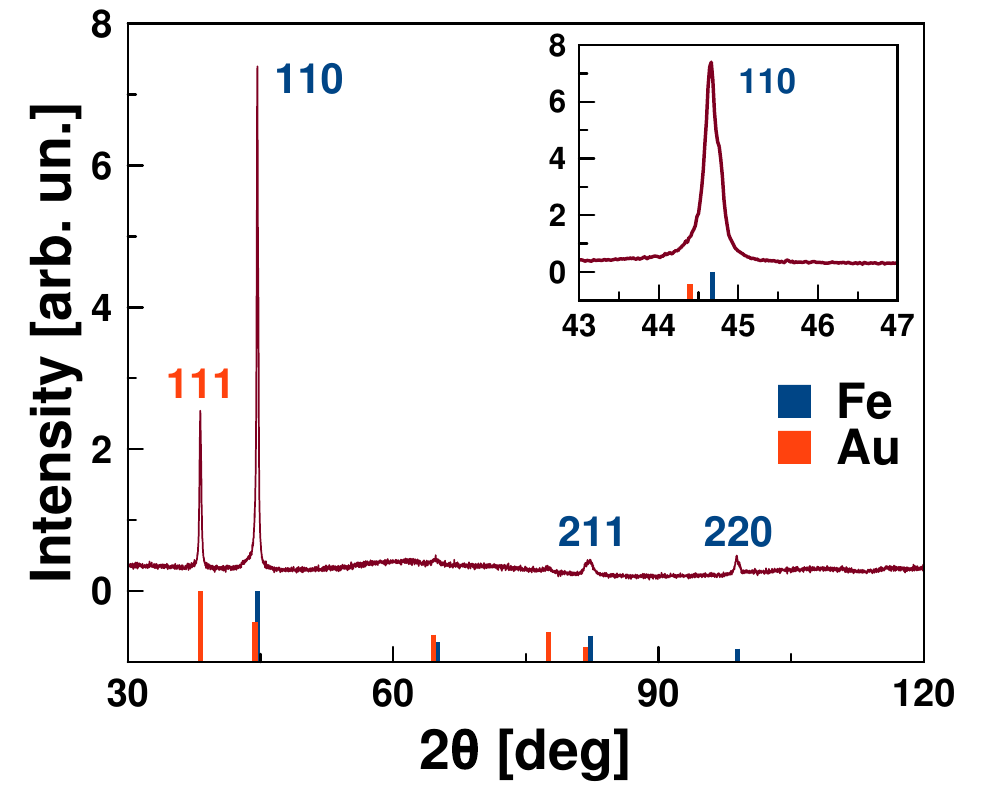}}
	\end{minipage}\\
	\caption{XRD pattern of Fe/AAO nanocomposite with 33~nm diameter Fe nanowires.}
	\label{ris:xrd}
\end{figure}

The phase composition of iron nanowires was proved by a Rigaku D/MAX 2500 X-ray diffractometer. The measurements were performed in the Bragg-Brentano geometry using  CuK$_{\alpha}$ radiation ($\lambda$~=~1.5418~\AA) in the 2$\theta$ range from 30 to 120 degrees. Prior to XRD measurements, the gold current collector was removed from the bottom side of the AAO by ion etching in order to minimize the intensity of gold peaks. It can be clearly seen that the main diffraction peaks correspond to the  $\alpha$-Fe phase (Fig.~\ref{ris:xrd}). There are no peaks of any iron oxides in the XRD pattern.  The peak at $2\theta = 38^{\circ}$ can be attributed to some gold islands, which remained after ion etching. According to XRD patterns, the intensities of the (211) and (220) reflections are lower than for the (110) reflex. Consequently, the iron nanowires are texturized; some of the crystallites grow along the [110] crystallographic direction. March–Dollase approach has estimated the degree of preferred orientation to be about 50\%~\cite{rodriguez2001introduction, zolotoyabko2009determination}.   

\subsection{SQUID-magnetometry}
\label{subsec:squid}

Magnetization measurements were carried using a Quantum Design MPMS-5S SQUID magnetometer at the Institute of Condensed Matter Physics (Braunschweig, Germany). The magnetization reversal curves were measured in the range of magnetic fields from --20 to 20~kOe with a step from 100~Oe to 1~kOe depending on the field range. All measurements were carried out at a temperature of 300~K.

\subsection{Analytical and micromagnetic models}
\label{subsec:mmc}


Interaction field and coercivity were first calculated by analytical models. An interaction field was considered as a maximal magnetic field produced by the nanowire array in a position of a particular nanowire. The nearest-neighboring nanowires were treated directly using a surface charges model while the outer ones are considered using mean field model. The field values were calculated at the assumed location of the nucleation volume center. The resulting field depends only on the diameter to structure period ratio $D/a$. The details are given elsewhere~\cite{da2011reduction}.

The value of the coercivity is strongly dependent on the nanowire reversal mode. If the remagnetization occurs by means of the movement of the TDW one can use the modification of the Stoner–Wohlfarth  model which takes into account domain wall width~\cite{vivas2012magnetic, lavin2009angular, hertel2015analytic, landeros2007reversal}. In the case of the VDW the coercivity of the nanowire can be calculated using the relation between vortex state length and external magnetic field. This can be found in turn by minimizing the vortex magnetic energy~
\cite{bochmann2018preparation, jamet2015head}. Then the coercivity can be defined as a field value at which the vortex length diverges. It seems reasonable to apply VDW model to all of the considered nanowires arrays since their diameter is higher than $7l_{exch}\approx25$ nm. However we have found out that TDW model describes coercivity of the thinnest nanowires much better than VDW one as will be shown below.  

If the field is applied perpendicular to the long axes of the nanowires remagnetization process can be considered as pseudo-coherent rotation. Most of the magnetic moments remagnetized coherently except the moments located at the caps of the nanowires (Fig.~\ref{ris:19_Hper}). Near saturation nanowires are almost uniformly magnetized and hence the saturation field can be calculated as $M_S(1 - p)/2$~\cite{bruck2018handbook}. Coercivity is negligibly small in this case.

Micromagnetic modeling was carried out by means of a numerical solution of the Landau-Lifshitz-Gilbert (LLG) equation using the finite element method~\cite{fredkin1990hybrid}. The calculations were performed in terms of the Nmag package provided by the University of Southampton~\cite{fischbacher2007systematic}. The following bulk parameters of bcc iron were used for modeling: exchange stiffness constant $A=2.1\cdot10^{-11}$~J/m and saturation magnetization $M_S=1.7\cdot 10^6$~A/m~\cite{bonilla2017magnetic}. The linear size of the finite element did not exceed the exchange length, which is 3.5~nm for iron. 
The magnetocrystalline anisotropy energy of iron ($K_1=5\cdot10^4$ J/m$^3$ )~\cite{cullity2011introduction}
 is relatively small compared to other terms in the full magnetic energy expression. Iron grains are mainy oriented along [110] crystallographic direction which is a intermediate axis for BCC iron. Therefore magnetocrystalline anisotropy only slightly changes the whole nanowire anisotropy~\cite{ivanov2015micromagnetic}. Taking all these facts into account we neglected anisotropy term in the first approximation. The length of the nanowires was chosen to be 400 nm during the simulation, which is more than an order of magnitude smaller than the length of the experimentally studied ones. Unfortunately, it is not possible to calculate the magnetization distribution in long (20-30 $\mu$m) nanowires using available resources. However, we have found that increase of the nanowire length to 700 nm does not affect the results of the simulations. We suggest that the aspect ratio of the nanowires is high enough to capture the most significant magnetic properties of the system. In addition, it should be noted that the results of the analytical and numerical model are close to each other for $D=52$ nm and $D=70$ nm  although the analytical model assumes that the nanowires are long (see Section~\ref{sec:results}).        

Calculations were performed for the arrays consisting of 7 and 19 nanowires. In the case of a 19 nanowire array a macrogeometry (MG) approach was used as well~\cite{fangohr2009new}. It allows one to create copies of the studied system, which exhibits exactly the same magnetization distribution, but their stray fields are taken into account. As a result it becomes possible to incorporate the effects of the sample size and shape into the model. Using this method, 100 copies of the 19 nanowire array were created and arranged in a hexagonal lattice. The magnetic field of these copies was taken into account when calculating the distribution of magnetization in the original system. 

We have also applied the simple mean field (MF) model to our system~\cite{zighem2011dipolar, panagiotopoulos2013packing}. We assumed that the mean field produced by the array can be calculated as $pM$, where $p$ is the porosity of the system. Strictly speaking the MF model can be applied to homogeneously magnetized systems only. Nevertheless, it is interesting to test its capabilities in the case of an iron nanowire array since it can substantially decrease the numerical effort.  

In sum, the following models were considered and compared:
\begin{enumerate}
	\item Analytical model 
	\item Micromagnetic simulation of 7 nanowires array
	\item Micromagnetic simulation of 7 nanowires array and MF model
	\item Micromagnetic simulation of 19 nanowires using MG approach
	\item Micromagnetic simulation of 19 nanowires using MG approach and MF model
\end{enumerate}

It should be noted that the last model is likely to significantly overestimate the value of the demagnetizing field. However, we have included it as the limit case. 

In all simulations and experiments both the parallel and perpendicular orientation of the external magnetic field with respect to the wires was considered.

\section{Results and discussion}
\label{sec:results}

\subsection{Magnetic field parallel to the long axes of the nanowires}
\label{subsec:outofplane}

\begin{figure*}[hptb]
	\begin{minipage}{0.55\linewidth}
		\center{\includegraphics[width=1\linewidth]{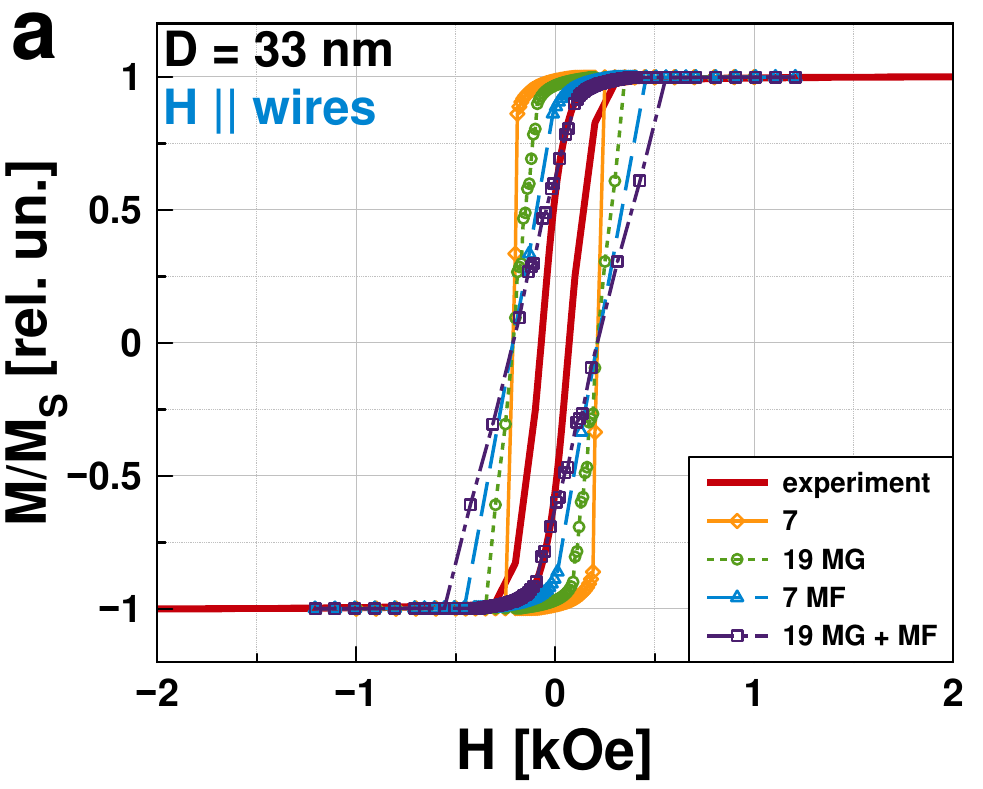}}
	\end{minipage}
    \begin{minipage}{0.55\linewidth}
		\center{\includegraphics[width=1\linewidth]{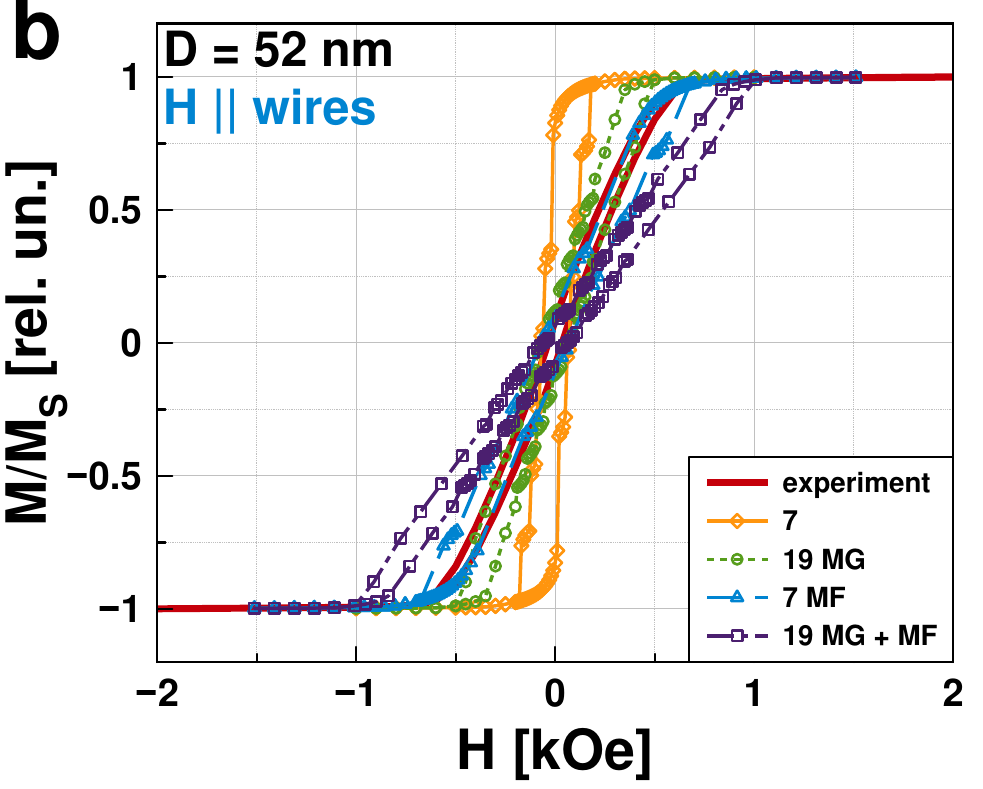}}
	\end{minipage}
	\begin{minipage}{0.55\linewidth}
		\center{\includegraphics[width=1\linewidth]{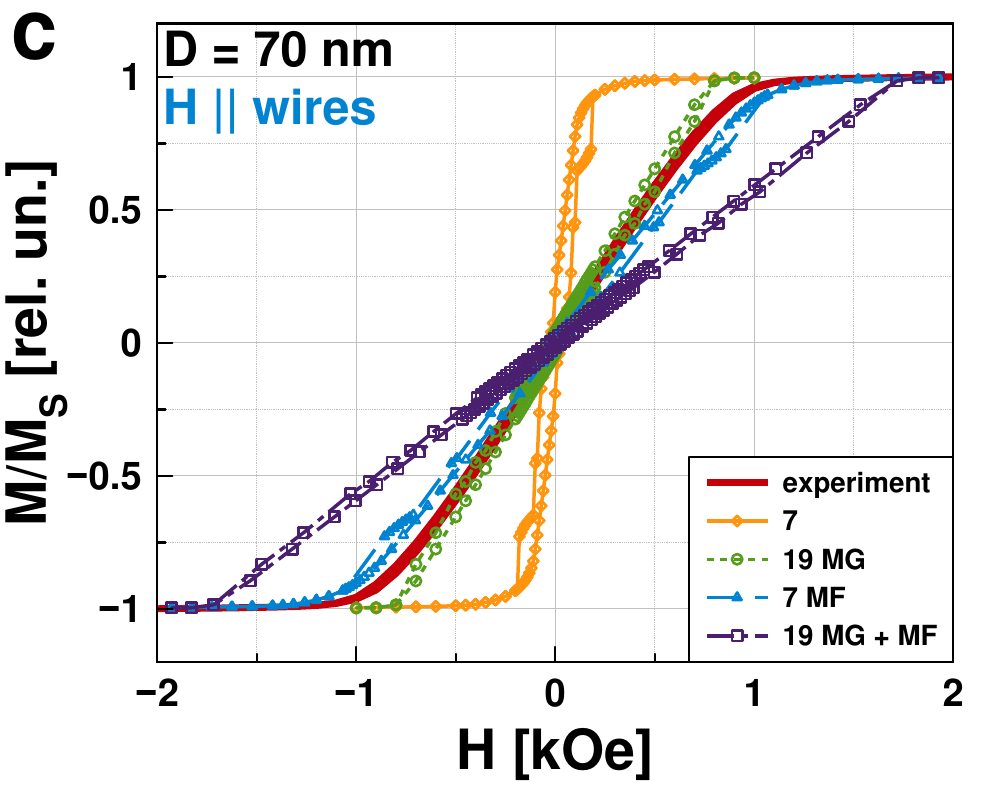}}
	\end{minipage}
	\caption{Hysteresis loops for the arrays of Fe nanowires with diameters of (a) 33,  (b) 52 and (c) 70~nm calculated using different models and measured experimentally. The magnetic field is applied parallel to the long axes of the nanowires.}
	\label{ris:diff_model}
\end{figure*}

\begin{figure}[hptb]
	\begin{minipage}{0.55\linewidth}
		\center{\includegraphics[width=1\linewidth]{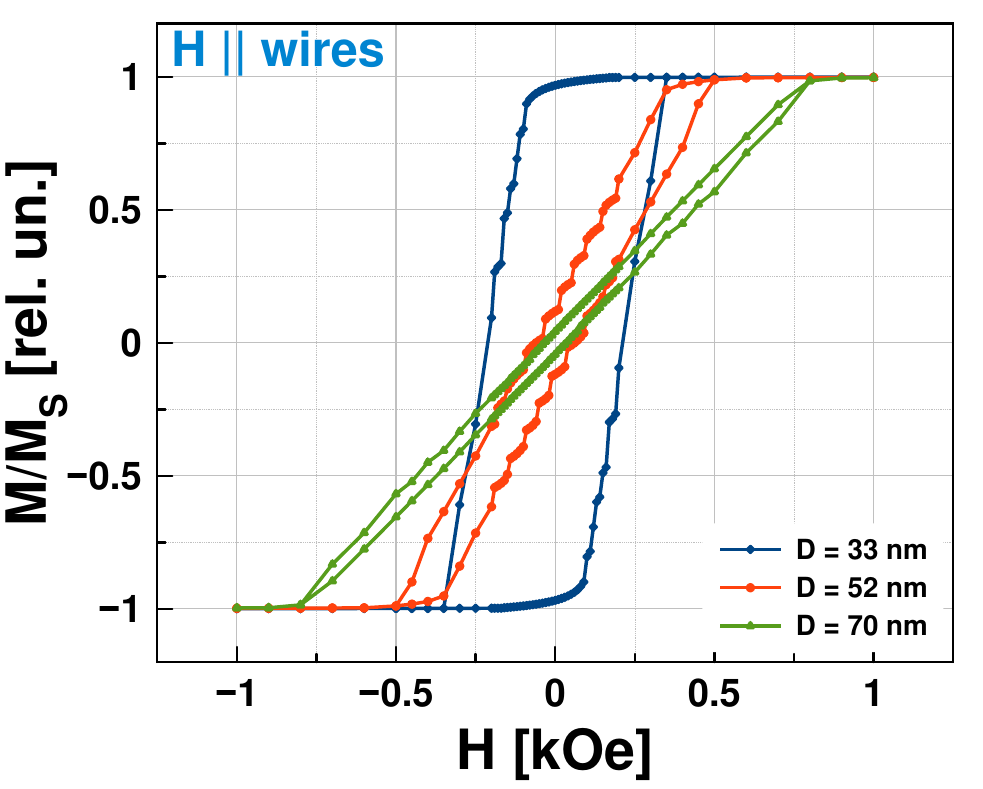}}
	\end{minipage}
   	\caption{Hysteresis curves for the arrays consisting of 19 nanowires with a diameter of 33, 52 and 70~nm, calculated using the macrogeometry model. The magnetic field is applied parallel to the long axes of the nanowires.}
	\label{ris:diameters_parall}
\end{figure}

The calculated and measured hysteresis loops are shown in Fig.~\ref{ris:diff_model}. It can be seen that an increase in the size of the system (number of nanowires) has the greatest effect on the behavior of large diameter nanowire arrays. (Fig.~\ref{ris:diff_model}). This is connected mainly to vortex states that appear in nanowires. These states decrease the nucleation fields of the nanowires which make them more sensitive to the stray fields produced by neighboring nanowires, which can be clearly seen in Fig.~\ref{ris:diameters_parall}.

\begin{figure*}[hptb]
	\begin{minipage}{0.33\linewidth}
		\center{\includegraphics[width=1\linewidth]{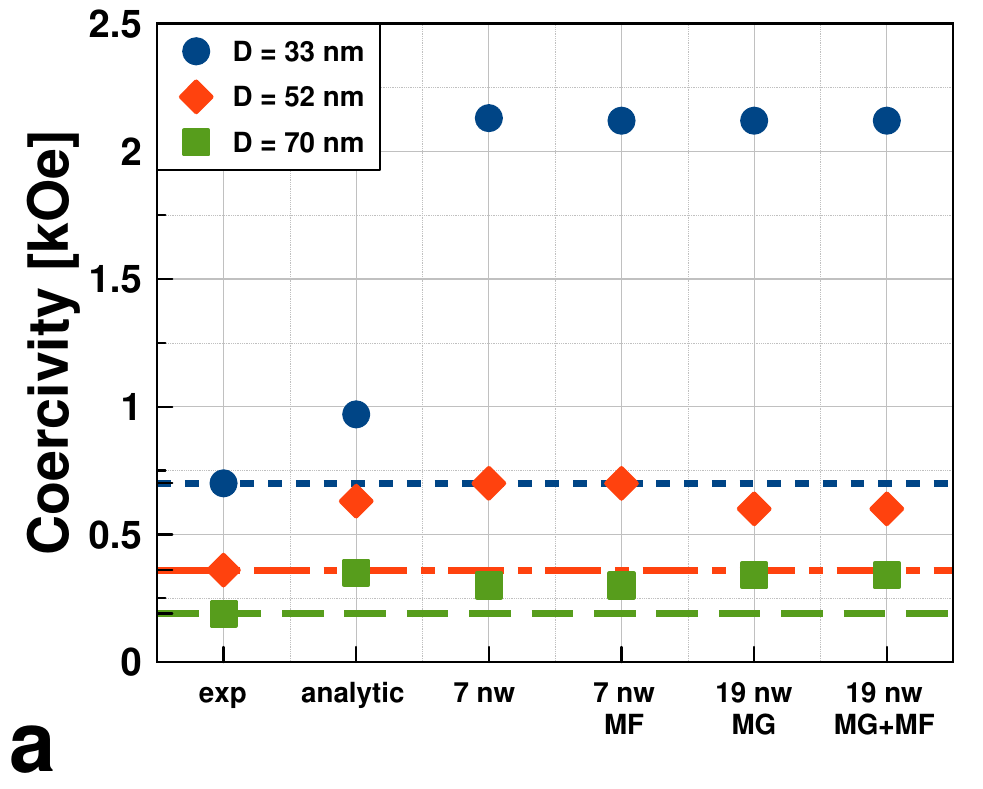}}
	\end{minipage}
    \begin{minipage}{0.33\linewidth}
		\center{\includegraphics[width=1\linewidth]{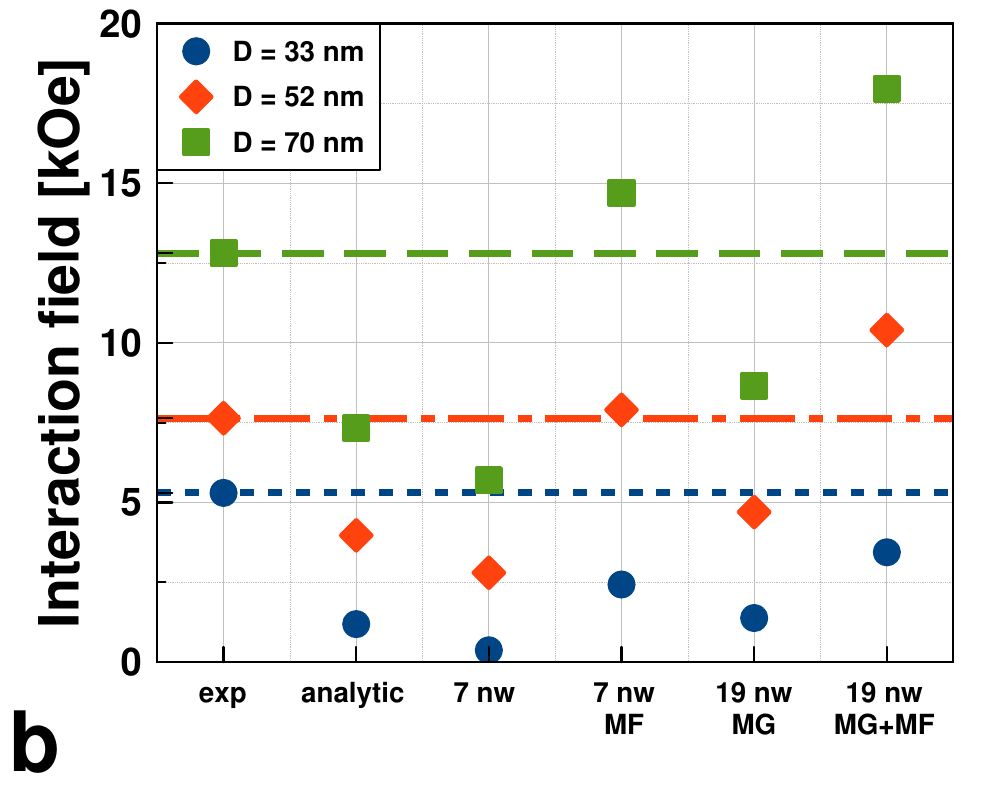}}
	\end{minipage}
	\begin{minipage}{0.33\linewidth}
		\center{\includegraphics[width=1\linewidth]{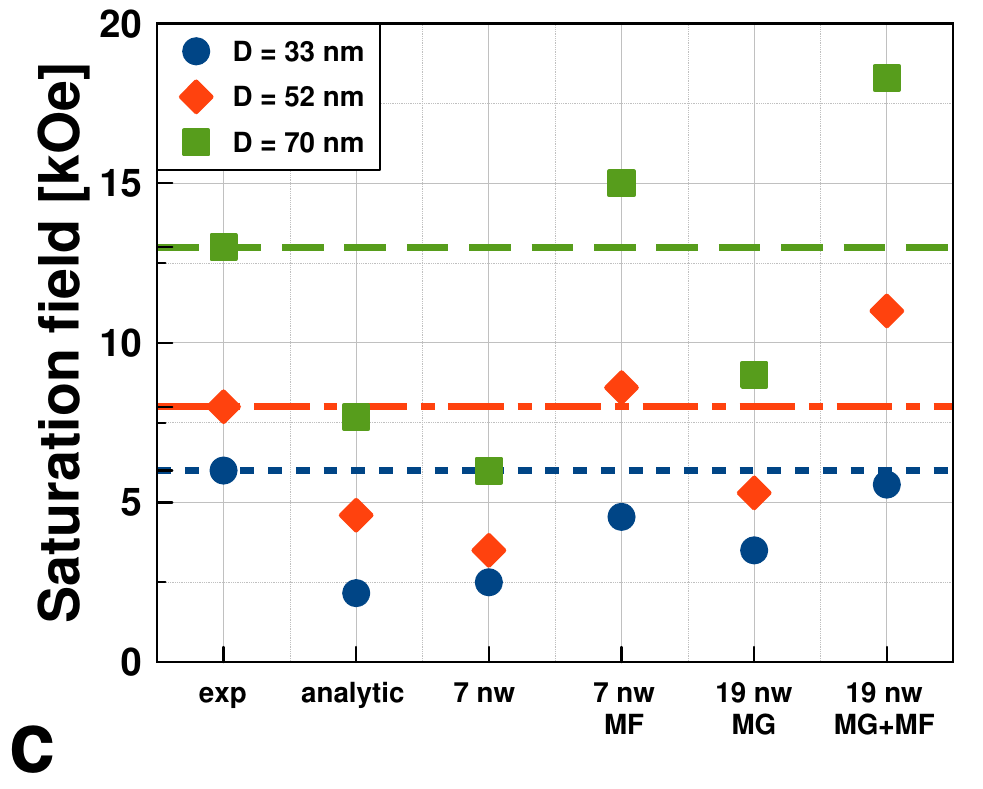}}
	\end{minipage}
	\caption{The results of calculations of (a) the coercivity, (b) the interaction field and (c) the saturation field for nanowires with a diameter of 33, 52, and 70~nm in the framework of different models and experimental data.}
	\label{ris:models}
\end{figure*}

\begin{figure*}[hptb]
	\begin{minipage}{0.33\linewidth}
		\center{\includegraphics[width=1\linewidth]{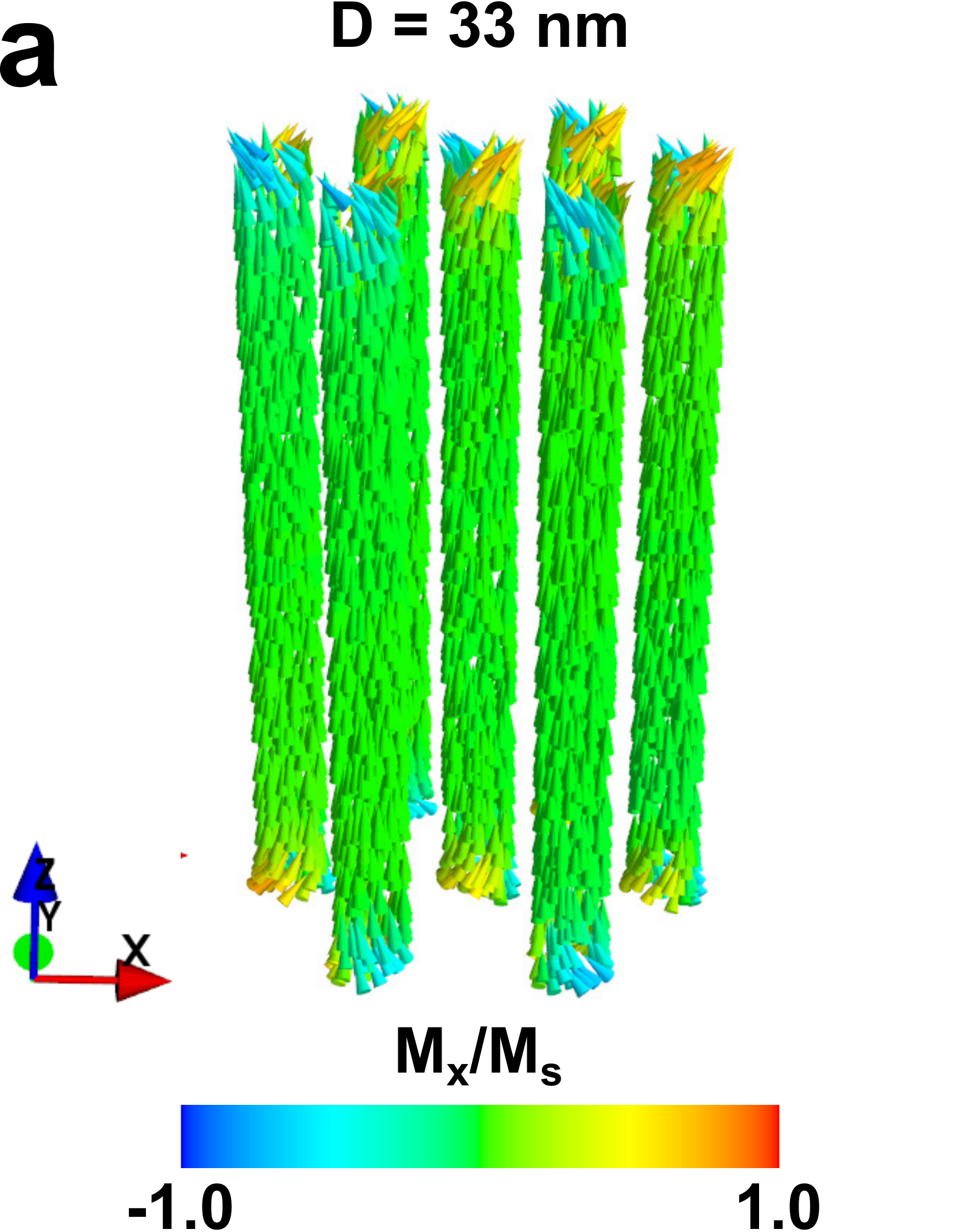}}
	\end{minipage}
    \begin{minipage}{0.33\linewidth}
		\center{\includegraphics[width=1\linewidth]{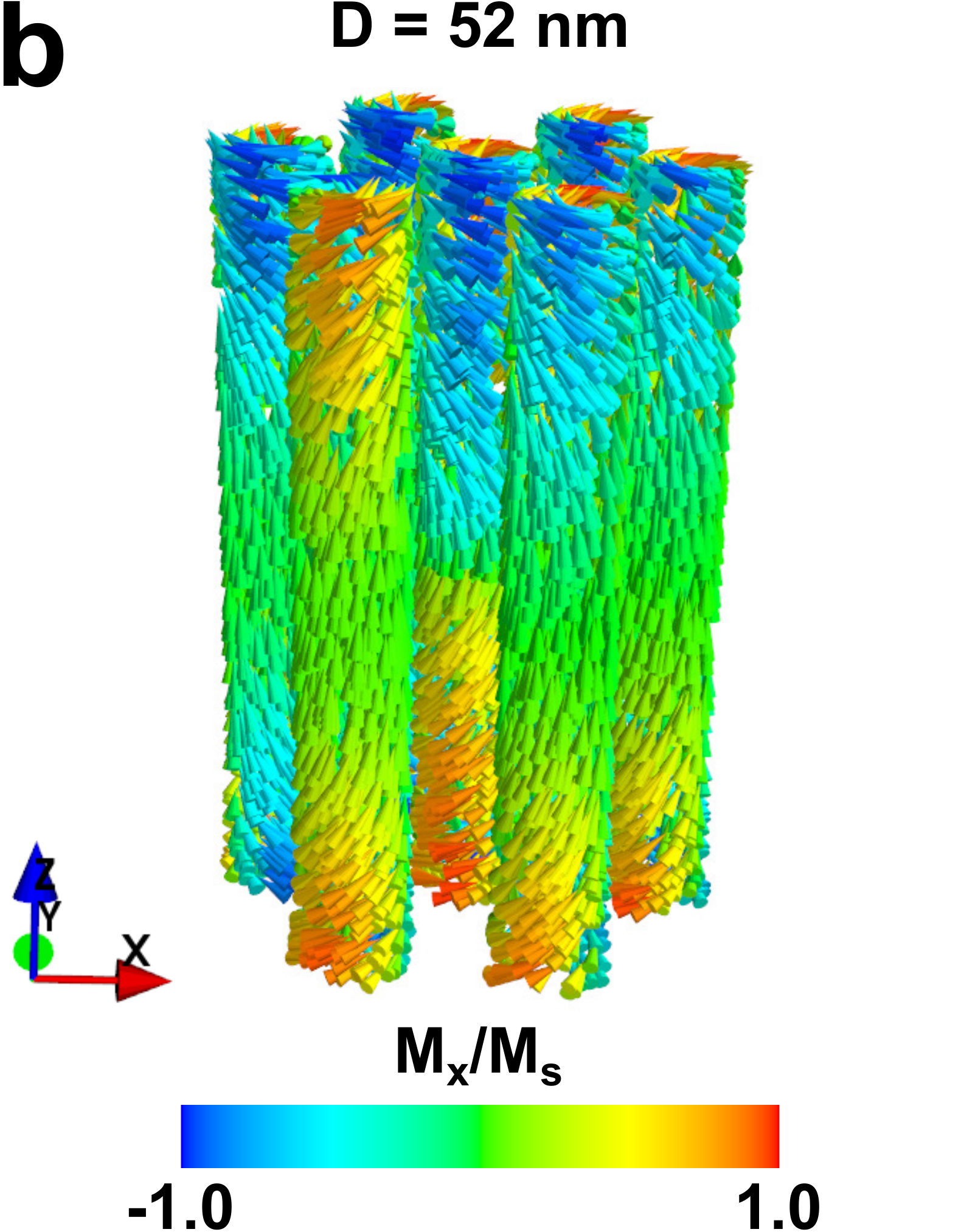}}
	\end{minipage}
	\begin{minipage}{0.33\linewidth}
		\center{\includegraphics[width=1\linewidth]{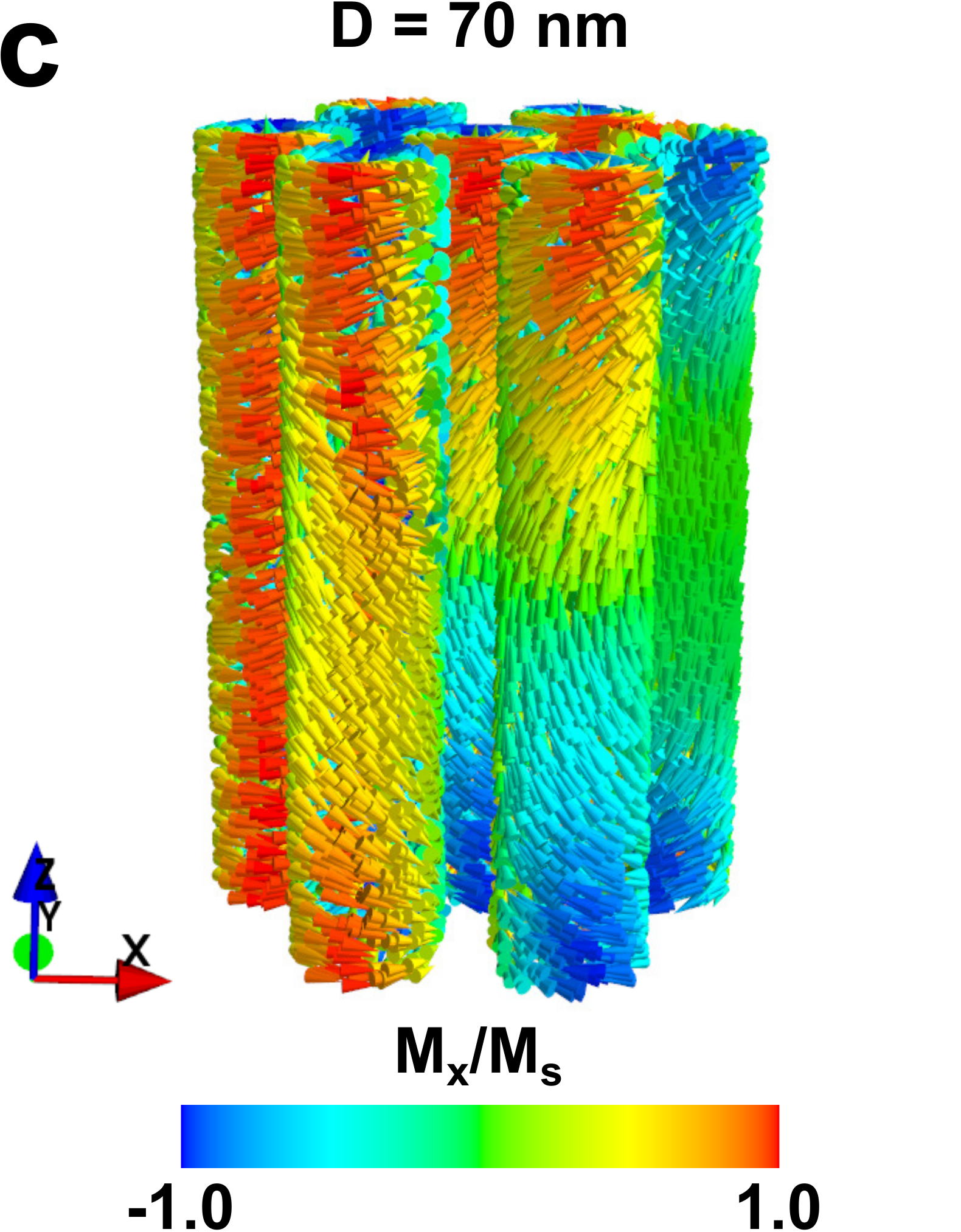}}
	\end{minipage}
	\caption{The distribution of magnetization in the remanence state after applying a saturation field along the axis of the nanowires. The diameters of nanowire are (a) 33~nm, (b) 52~nm, and (c) 70~nm. The color indicates the magnitude of the projection of the magnetization on the horizontal oX axis. The oZ axis coincides with the long axes of the nanowires.}
	\label{ris:7_Hpar}
\end{figure*}

Coercivity data are shown in Fig.~\ref{ris:models}(a).
In the case of large diameter nanowires (52 and 70 nm) both analytical and micromagnetic models predict coercivity values that are close to the experimental ones. Here we have used model that describes VDW.  One can see that the vortex occupies a significant volume of the nanowires with a diameter of 52 nm and almost the entire volume of those with a diameter of 70 nanometers (Fig.~\ref{ris:7_Hpar}). The vortex state almost completely determines the magnetization reversal mode of the nanowire in these cases.

However the situation changes in the case of the wires with a diameter of 33~nm. Analytical model which is based on the assumption that the vortex state has a major effect on the magnetization process~\cite{bochmann2018preparation} predicts the overestimated coercivity of 1.6 kOe. TDW model~\cite{lavin2009angular} gives the value of 0.97 kOe which is closer to the measured one (0.7 kOe) (Fig.~\ref{ris:models}(a)). Moreover micromagnetc calculations show the absence of the vortex states in 33 nm nanowires. Transverse end domains arise in the nanowires caps instead (Fig.~\ref{ris:7_Hpar}(a)). Therefore one can conclude that the magnetic behavior of these wires is not determined by vortex states, despite the fact that the diameter of the wires exceeds $7l_{exch}$. It should be noted that micromagnetic simulations hugely overestimate the value of the coercivity. The reason may be that, in the framework of the analytical model, wires are considered long while the aspect ratio of the numerically simulated wires is not large enough. Introducing small distortions into the wires shape may significantly decrease coercivity and brings it closer to the experimental values~\cite{panagiotopoulos2015athermal}. However such approach requires a large number of finite elements and is hardly tractable for materials with small exchange length value. As expected, the differences between the various micromagnetic models are small since they are mainly focused on taking into account interactions between nanowires in different ways.           

Interaction field values are presented in Fig.~\ref{ris:models}(b). The analytical model does not agree well with the experimental data. The minimal difference is observed for nanowires of 52 nm in diameter. It is possible that for thin nanowires structural defects may alter the interactions, while for the thick ones the analytical model does not quite accurately take into account the large vortex states which occupy a significant volume of the nanowires. Micromagnetic models applied to the array of 33 nm nanowires lead to the values of the interaction field that also differ from the experimental ones. Perhaps in this case the length of the nanowires as well as structural inhomogeneities play a role.

However, the predictive capabilities of the numerical models increase with the growth of nanowires diameters.  In the case of Fe$_{52}$ nanowires the calculated value of the interaction field (7.9 kOe) is very close to the measured one (7.6 kOe). For Fe$_{70}$ sample the difference is larger but still acceptable: 12.8 kOe and 14.7 kOe for the simulation and experiment respectively. Surprisingly, the MF approach for 7 nanowires predicts the values of the interaction fields better than the MG model for 19 nanowires. This may be due to the fact that the arrays of nanowires investigated experimentally consist of randomly oriented structural domains with a characteristic length of about 1 $\mu$m or (approximately 10 lattice periods). Therefore, taking into account only the nearest neighbors is more accurate than calculations made for 19 ordered nanowires by means of MG approach, which assumes that the ideally ordered cluster consists of about 30000 wires. At the same time, a model which takes into account only 7 nanowires requires significantly less computational resources.

The saturation field is defined as a field at which the magnetization exceeds 0.99, or as a field in which the angle of inclination of the hysteresis loop has significantly changed for the last time. The highest value is selected. The analytical model is in the best agreement with the results for 19 nanowires obtained using the macrogeometry model (Fig.~\ref{ris:models}(c)). Again we have obtained a good match between simulations in the frame of MF model and the experiments for Fe$_{52}$ and Fe$_{70}$ samples.

\subsection{Magnetic field perpendicular to the long axes of the nanowires}
\label{subsec:inplane}

\begin{figure*}[hptb]
	\begin{minipage}{0.33\linewidth}
		\center{\includegraphics[width=1\linewidth]{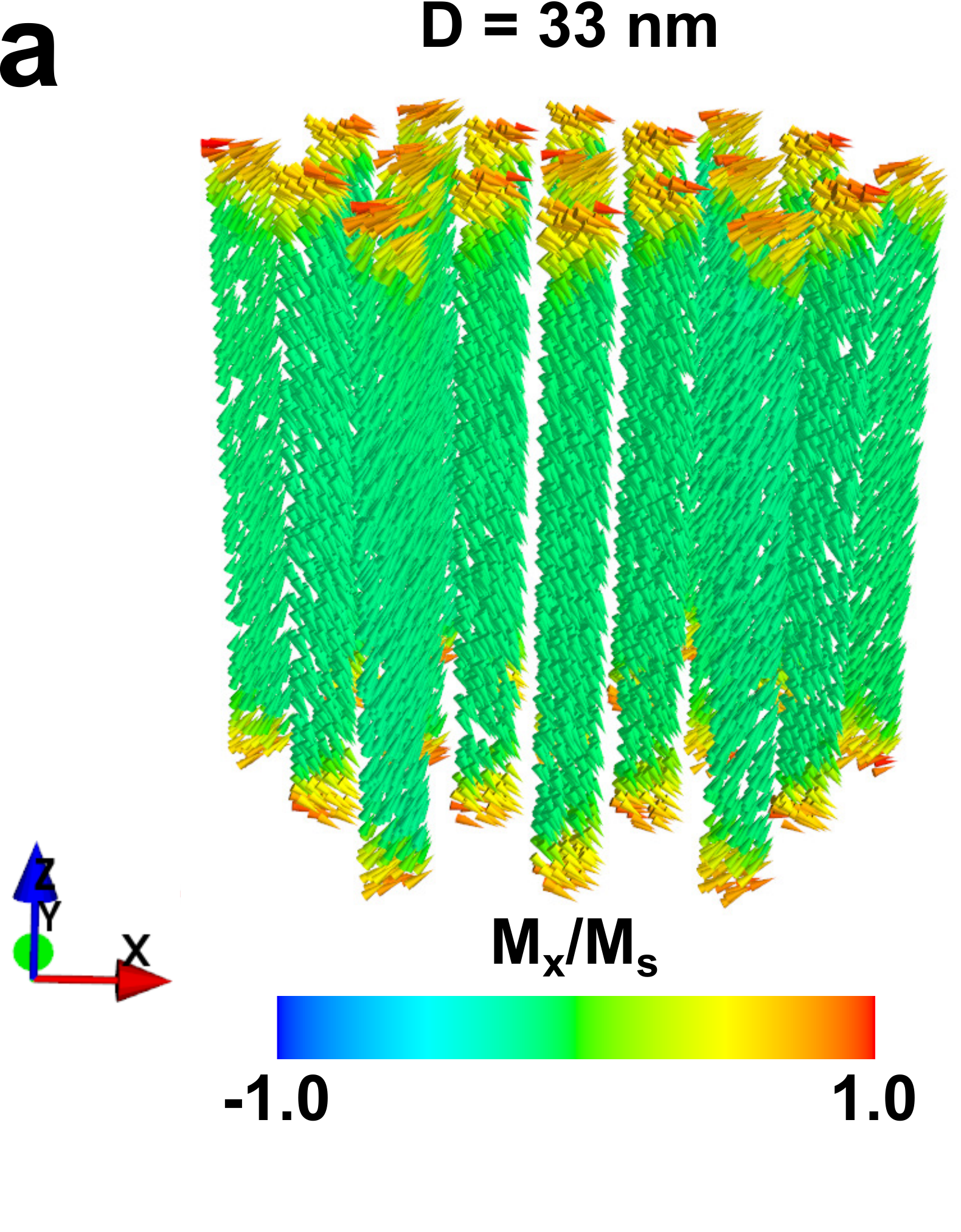}}
	\end{minipage}
    \begin{minipage}{0.33\linewidth}
		\center{\includegraphics[width=1\linewidth]{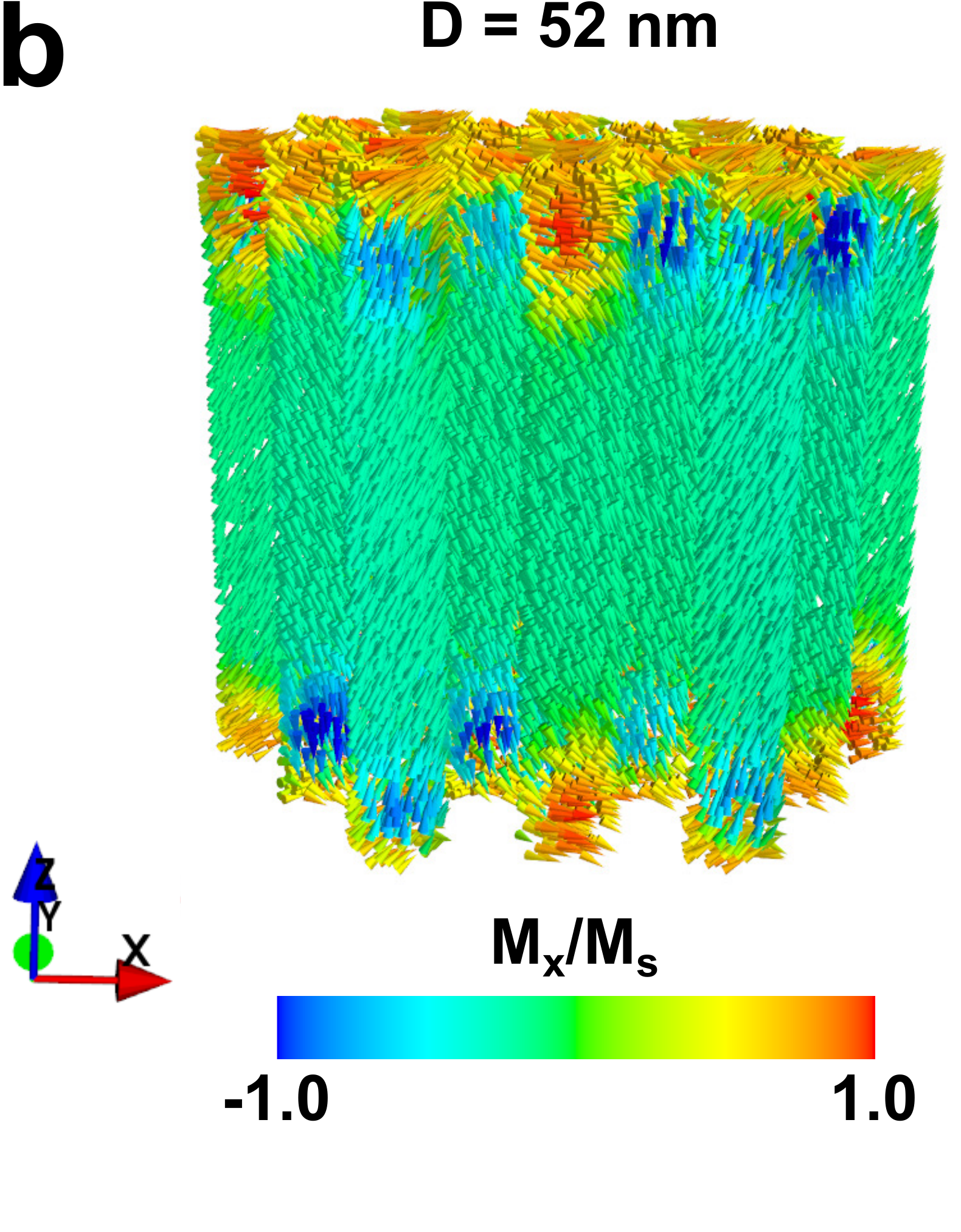}}
	\end{minipage}
	\begin{minipage}{0.33\linewidth}
		\center{\includegraphics[width=1\linewidth]{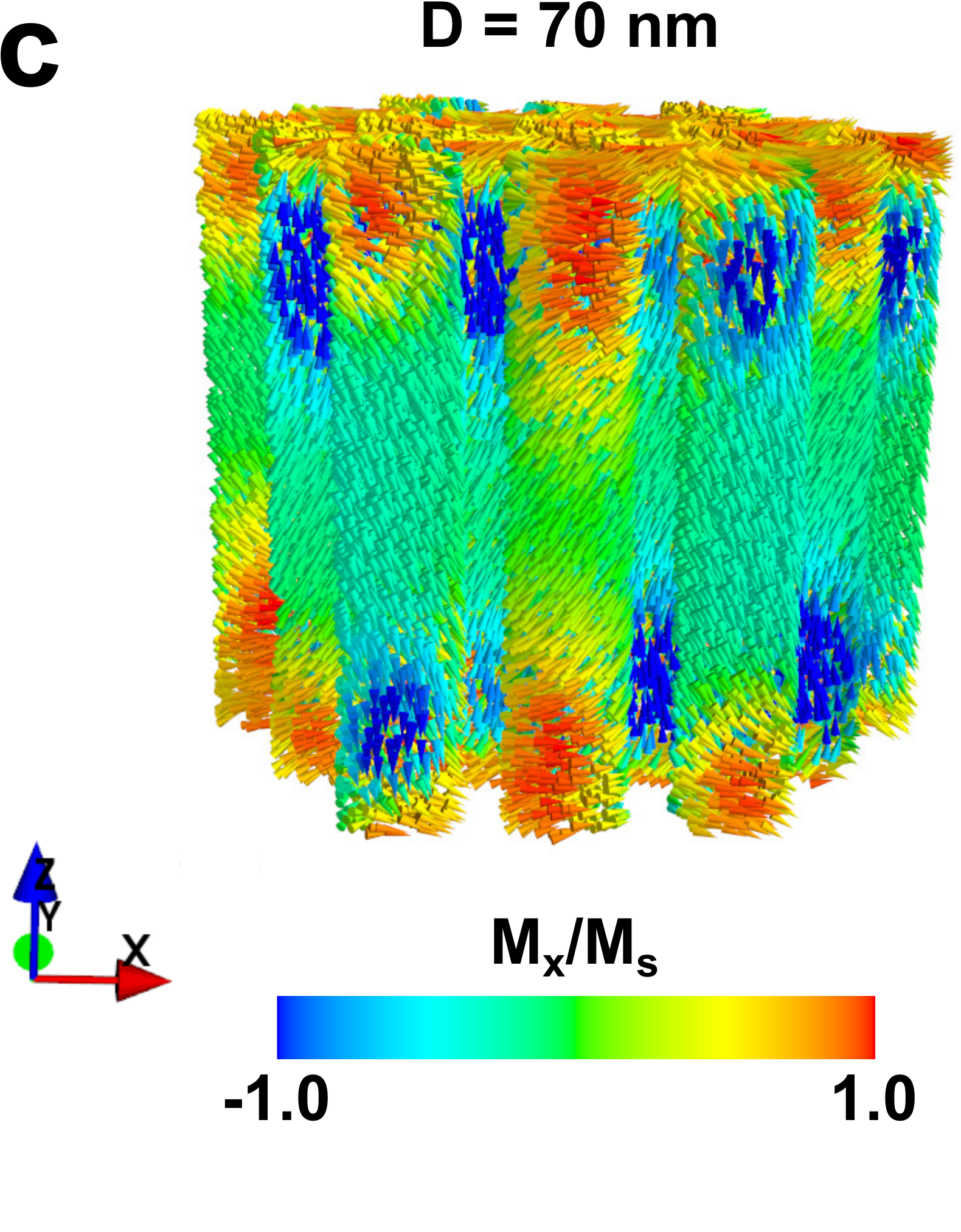}}
	\end{minipage}
	\caption{The distribution of magnetization in an array consisting of 19 nanowires in a field applied perpendicular to the long axes of the nanowires. The magnetization m$_x$~=~0.5, the field decreases. The diameters of nanowires are 33, 52 and 70~nm. The calculation was carried out without using the MG model.}
	\label{ris:19_Hper}
\end{figure*}

\begin{figure}[hptb]
	\begin{minipage}{0.55\linewidth}
		\center{\includegraphics[width=1\linewidth]{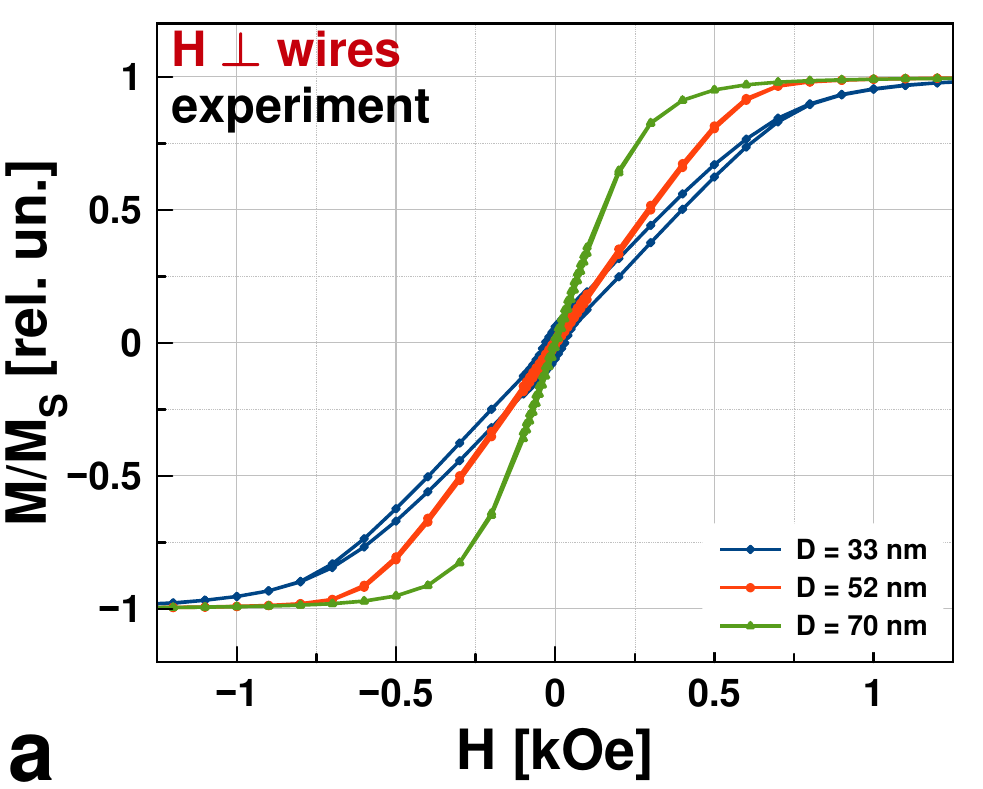}}
	\end{minipage}\begin{minipage}{0.55\linewidth}
		\center{\includegraphics[width=1\linewidth]{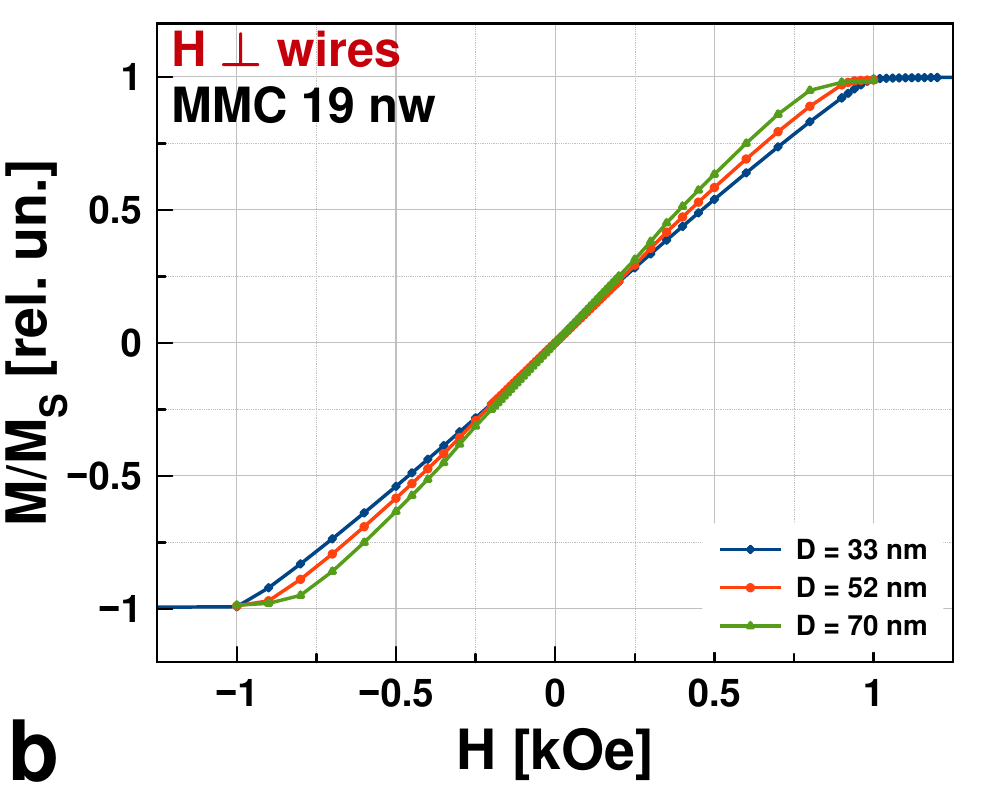}}
	\end{minipage}
	\caption{Hysteresis loops for the arrays of Fe nanowires with the diameters of 33, 52 and 70 nm: experimental (a) and calculated by means of micromagnetism for 19 nanowires (b). The magnetic field is applied perpendicular to the long axes of the nanowires.}
	\label{ris:diameters_perp}
\end{figure}

In the case of perpendicular orientation of the external magnetic field to the long axes of the nanowires, the values of coercivity is negligibly small, so only the saturation field has been analyzed. Significant deviation between the models and the experiments is observed for the 33 nm nanowires (Table~\ref{tab:Hper_data_33}). Similar to the case of the parallel direction of the magnetic field, the possible reason of such difference may be defects in the nanowire structure and the large lengths of the nanowires studied experimentally. As expected, the saturation field for the thickest nanowires shows a better agreement to the micromagnetic simulation, since the analytical model is based on the assumption of coherent rotation of the magnetization, whereas vortex states arise in large diameter nanowires (Fig.~\ref{ris:19_Hper}). However, it should be noted that both the numerical and analytical models suggest that the saturation field should decrease monotonically with increasing diameter. Nevertheless, there is a nonmonotonic dependence in the experiment.  We suggest that this dependence may be caused by different pore filling factor of the considered samples~\cite{goncharova2017oriented}. Pore filling factor $f$ may decrease up to 50\% and hence alter the porosity. The decrease of the effective porosity $pf$ leads to the increase of the saturation field which may be the case for Fe$_{70}$ sample. However, the calculated and measured hysteresis curves look qualitatively similar, as can be seen in Figure~\ref{ris:diameters_perp}.  

\begin{table}[hbtp]
    \centering
    \resizebox{\columnwidth}{!}{
    \begin{tabular}{l|c|c|c}
        & \multicolumn{3}{c}{Saturation field, kOe} \\
        \hline
        Approach & $D=33$~nm & $D=52$~nm & $D=70$~nm \\
        
       \hline
       Experiment & 15.0~$\pm$~0.5  & 8.0~$\pm$~0.5 & 10.0~$\pm$~0.5 \\
       Analytical model & 9.7 & 8.2 & 6.1 \\
       Micromagnetic calculation (19 nanowires) & 10.0 & 9.4 & 9.0
    \end{tabular}
    }
    \caption{Values of the saturation field obtained using different approaches; the external field is applied perpendicular to long axes of the nanowires.}
    \label{tab:Hper_data_33}
\end{table}

\begin{figure}[hptb]
	\begin{minipage}{0.55\linewidth}
		\center{\includegraphics[width=1\linewidth]{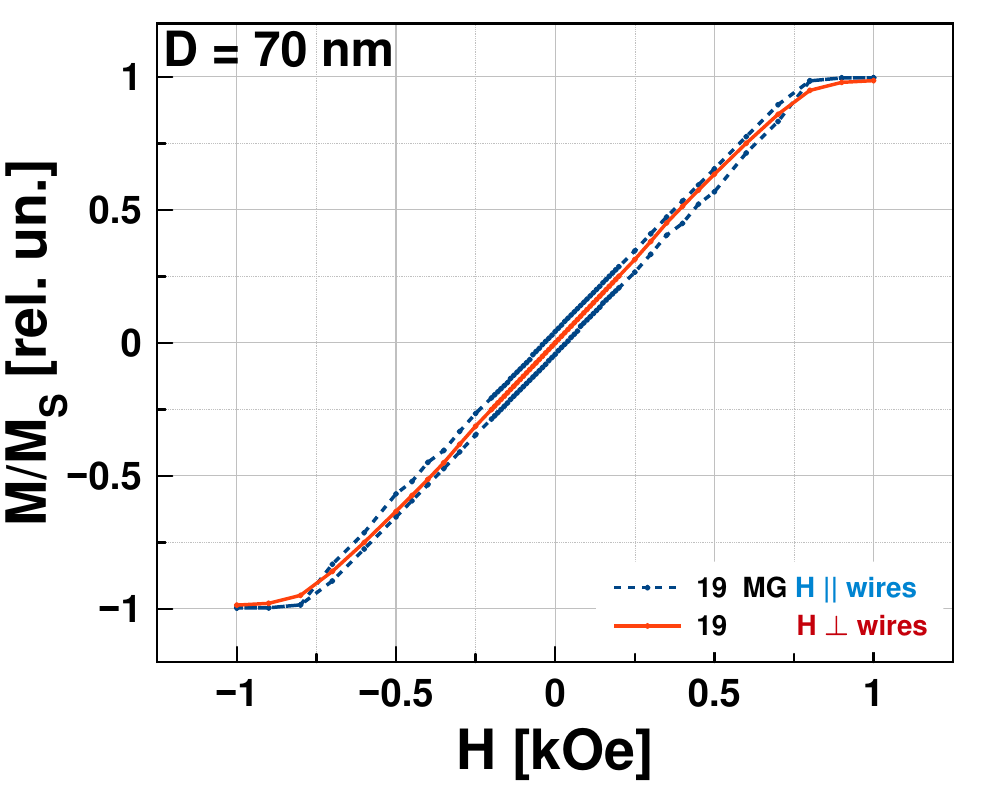}}
	\end{minipage}
   	\caption{Magnetization curves for a system consisting of 19 nanowires (the field is perpendicular to the axis of the nanowires) and for a system of 19 nanowires, calculated using the MG model (the field is parallel to the axis of the nanowires), D~=~70~nm}
	\label{ris:two_geometries}
\end{figure}

Interestingly, the slopes of the magnetization reversal curves calculated for different directions of the external magnetic field are similar to each other for a system of nanowires with a diameter of 70 nm (Fig.~\ref{ris:two_geometries}). A similar behavior was observed in Ref~\cite{dobosz2019synthesis} for 3 $\mu$m lengths iron nanowires. Hence vortex states almost completely suppress shape anisotropy of short nanowires and make them isotropic.

\section{Concluding remarks}
\label{sec:conclusion}

To conclude, we have studied three arrays of iron nanowires with different diameters. Several micromagnetic models of the magnetization reversal and analytical theory have been compared with the experiment. The key parameters that describe the hysteresis loops of the nanowire arrays are the interaction field and the coercivity. The former is connected mainly to the interactions of the nanowires, whereas the latter is influenced mostly by their individual switching field. Thus, by analyzing these parameters, it is possible to understand the predictive capability of the theoretical models. It was shown that vortex states have a significant influence on the behavior of the nanowires with diameters of 52 and 70~nm. We have found out that in the case of the array of iron nanowires which consists of randomly oriented hexagonal regions with an average size of about 1$\mu$m, the best agreement with the experiment is reached when applying the micromagnetic model which takes into account 7 nanowires in the mean field approximation. Numerical models agree better with the experimental data for the thick nanowires. This may be due to the fact that their behavior is mostly determined by the vortex state and weakly depends on the length of the nanowires and the quality of their structure. Therefore, more complex magnetic states provide a better agreement between the models and the experiments.

\section{Acknowledgements}
\label{sec:acknowlegdements}
This work was supported by Russian Science Foundation (project No. 18-72-00011). The authors are thankful to the Research Park of St. Petersburg State University and the staff of the resource centers “Nanotechnologies” (http://nano.spbu.ru) and “X-ray diffraction methods of research” (http://xrd.spbu.ru) V. Kalganov and I. Kasatkin, respectively, for their invaluable contribution in obtaining experimental data. We are grateful to G. A. Valkovskiy for the help with the analysis of XRD measurements.  The computing resources of the Resource Center “Computing Center of St. Petersburg State University” (http://www.cc.spbu.ru/) were used for carrying out the micromagnetic calculations. A part of them was performed with the help of the data center PIK NRC “Kurchatov Institute” – PNPI. In addition, the authors thank the developers of the Magic Plot software (https://magicplot.com/), in which the most part of the data was processed and plotted. The authors declare that there is no conflict of interest regarding the publication of this article.

\section*{References}

\bibliographystyle{unsrt}
\bibliography{biblio}

\begin{thebibliography}{10}

\bibitem{DEMIREL2015104}
S~Demirel, E~Oz, E~Altin, S~Altin, A~Bayri, P~Kaya, S~Turan, and S~Avci.
\newblock Growth mechanism and magnetic and electrochemical properties of
  na0.44mno2 nanorods as cathode material for na-ion batteries.
\newblock {\em Materials Characterization}, 105:104--112, 2015.

\bibitem{supercond}
F~Hekmat, S~Shahrokhian, and S~Rahimi.
\newblock 3d flower-like binary nickel cobalt oxide decorated coiled carbon
  nanotubes directly grown on nickel nanocones and binder-free hydrothermal
  carbons for advanced asymmetric supercapacitors.
\newblock {\em Nanoscale}, 11(6):2901--2915, 2019.

\bibitem{SU20091062}
Y~Su, M~Gao, X~Meng, Y~Chen, Q~Zhou, L~Li, and Y~Feng.
\newblock Synthesis of in-doped ga2o3 zigzag-shaped nanowires and optical
  properties.
\newblock {\em Journal of Physics and Chemistry of Solids}, 70(7):1062--1065,
  2009.

\bibitem{Wierzbicki_2015}
M~Wierzbicki, J~Barna{\'s}, and R~Swirkowicz.
\newblock Zigzag nanoribbons of two-dimensional silicene-like crystals:
  magnetic, topological and thermoelectric properties.
\newblock {\em Journal of Physics: Condensed Matter}, 27(48):485301, 2015.

\bibitem{parkin2008magnetic}
S~Parkin, M~Hayashi, and L~Thomas.
\newblock Magnetic domain-wall racetrack memory.
\newblock {\em Science}, 320(5873):190--194, 2008.

\bibitem{bruck2018handbook}
Michal Sta{\v{n}}o and Olivier Fruchart.
\newblock Magnetic nanowires and nanotubes.
\newblock In {\em Handbook of Magnetic Materials}, volume~27, pages 155--267.
  Elsevier, 2018.

\bibitem{vazquez2015magnetic}
M~V{\'a}zquez.
\newblock {\em Magnetic nano-and microwires}.
\newblock Elsevier, 2015.

\bibitem{zeng2002structure}
H~Zeng, R~Skomski, L~Menon, Y~Liu, S~Bandyopadhyay, and D~J Sellmyer.
\newblock Structure and magnetic properties of ferromagnetic nanowires in
  self-assembled arrays.
\newblock {\em Physical Review B}, 65(13):134426, 2002.

\bibitem{ortega2017magnetic}
E~Ortega, S~M Reddy, I~Betancourt, S~Roughani, B~JH Stadler, and A~Ponce.
\newblock Magnetic ordering in 45 nm-diameter multisegmented fega/cu nanowires:
  Single nanowires and arrays.
\newblock {\em Journal of Materials Chemistry C}, 5(30):7546--7552, 2017.

\bibitem{ishii1989magnetic}
Y~Ishii and M~Sato.
\newblock Magnetic behavior of a film with columnar structure.
\newblock {\em Journal of magnetism and magnetic materials}, 82(2-3):309--312,
  1989.

\bibitem{sun2005tuning}
L~Sun, Y~Hao, C-L Chien, and P~C Searson.
\newblock Tuning the properties of magnetic nanowires.
\newblock {\em IBM Journal of Research and Development}, 49(1):79--102, 2005.

\bibitem{goncharova2017oriented}
A~S Goncharova, S~V Sotnichuk, A~S Semisalova, T~Y Kiseleva, I~Sergueev,
  M~Herlitschke, K~S Napolskii, and A~A Eliseev.
\newblock Oriented arrays of iron nanowires: synthesis, structural and magnetic
  aspects.
\newblock {\em Journal of Sol-Gel Science and Technology}, 81(2):327--332,
  2017.

\bibitem{schlesinger2011modern}
M~Schlesinger and M~Paunovic.
\newblock {\em Modern electroplating}, volume~55.
\newblock John Wiley \& Sons, 2011.

\bibitem{ivanov2013magnetic}
Y~P Ivanov, M~V{\'a}zquez, and O~Chubykalo-Fesenko.
\newblock Magnetic reversal modes in cylindrical nanowires.
\newblock {\em Journal of Physics D: Applied Physics}, 46(48):485001, 2013.

\bibitem{schaefer2016nico}
S~Schaefer, E-M Felix, F~Muench, M~Antoni, C~Lohaus, J~Br{\"o}tz, U~Kunz,
  I~G{\"a}rtner, and W~Ensinger.
\newblock Nico nanotubes plated on pd seeds as a designed magnetically
  recollectable catalyst with high noble metal utilisation.
\newblock {\em RSC advances}, 6(74):70033--70039, 2016.

\bibitem{pitzschel2011magnetic}
K~Pitzschel, J~Bachmann, S~Martens, J~M Montero-Moreno, J~Kimling, G~Meier,
  J~Escrig, K~Nielsch, and D~G{\"o}rlitz.
\newblock Magnetic reversal of cylindrical nickel nanowires with modulated
  diameters.
\newblock {\em Journal of Applied Physics}, 109(3):033907, 2011.

\bibitem{ruiz2019tailoring}
A~Ruiz-Clavijo, S~Ruiz-Gomez, O~Caballero-Calero, L~Perez, and
  M~Martin-Gonzalez.
\newblock Tailoring magnetic anisotropy at will in 3d interconnected nanowire
  networks.
\newblock {\em physica status solidi (RRL)--Rapid Research Letters},
  13(10):1900263, 2019.

\bibitem{bran2018magnetization}
C~Bran, E~Berganza, J~A Fernandez-Roldan, E~M Palmero, J~Meier, E~Calle,
  M~Jaafar, M~Foerster, L~Aballe, A~Fraile~Rodriguez, et~al.
\newblock Magnetization ratchet in cylindrical nanowires.
\newblock {\em ACS nano}, 12(6):5932--5939, 2018.

\bibitem{wolf2019holographic}
D~Wolf, N~Biziere, S~Sturm, D~Reyes, T~Wade, T~Niermann, J~Krehl,
  B~Warot-Fonrose, B~B{\"u}chner, E~Snoeck, et~al.
\newblock Holographic vector field electron tomography of three-dimensional
  nanomagnets.
\newblock {\em Communications Physics}, 2(1):87, 2019.

\bibitem{biziere2013imaging}
N~Biziere, C~Gatel, R~Lassalle-Balier, M~C Clochard, J~E Wegrowe, and E~Snoeck.
\newblock Imaging the fine structure of a magnetic domain wall in a ni
  nanocylinder.
\newblock {\em Nano letters}, 13(5):2053--2057, 2013.

\bibitem{forster2002domain}
H~Forster, T~Schrefl, D~Suess, W~Scholz, V~Tsiantos, R~Dittrich, and J~Fidler.
\newblock Domain wall motion in nanowires using moving grids.
\newblock {\em Journal of applied physics}, 91(10):6914--6919, 2002.

\bibitem{wieser2004domain}
R~Wieser, U~Nowak, and K~D Usadel.
\newblock Domain wall mobility in nanowires: Transverse versus vortex walls.
\newblock {\em Physical Review B}, 69(6):064401, 2004.

\bibitem{thiaville2006domain}
A~Thiaville and Y~Nakatani.
\newblock Domain-wall dynamics in nanowires and nanostrips.
\newblock In {\em Spin dynamics in confined magnetic structures III}, pages
  161--205. Springer, 2006.

\bibitem{ebels2000spin}
U~Ebels, A~Radulescu, Y~Henry, Luc Piraux, and K~Ounadjela.
\newblock Spin accumulation and domain wall magnetoresistance in 35 nm co
  wires.
\newblock {\em Physical review letters}, 84(5):983, 2000.

\bibitem{da2016nucleation}
S~Da~Col, S~Jamet, M~Sta{\v{n}}o, B~Trapp, S~Le~Denmat, L~Cagnon, J~C
  Toussaint, and O~Fruchart.
\newblock Nucleation, imaging, and motion of magnetic domain walls in
  cylindrical nanowires.
\newblock {\em Applied Physics Letters}, 109(6):062406, 2016.

\bibitem{stavno2017probing}
M~Sta{\v{n}}o, S~Jamet, J~C Toussaint, S~Bochmann, J~Bachmann, A~Masseboeuf,
  C~Gatel, and O~Fruchart.
\newblock Probing domain walls in cylindrical magnetic nanowires with electron
  holography.
\newblock In {\em Journal of Physics: Conference Series}, volume 903, page
  012055. IOP Publishing, 2017.

\bibitem{thiaville2003micromagnetic}
A~Thiaville, J~M Garc{\'\i}a, J~Miltat, T~Schrefl, et~al.
\newblock Micromagnetic study of bloch-point-mediated vortex core reversal.
\newblock {\em Physical Review B}, 67(9):094410, 2003.

\bibitem{fangohr2009new}
H~Fangohr, G~Bordignon, M~Franchin, A~Knittel, P~AJ de~Groot, and
  T~Fischbacher.
\newblock A new approach to (quasi) periodic boundary conditions in
  micromagnetics: The macrogeometry.
\newblock {\em Journal of Applied Physics}, 105(7):07D529, 2009.

\bibitem{vock2017role}
S~Vock, C~Hengst, Z~Sasv{\'a}ri, R~Sch{\"a}fer, L~Schultz, and V~Neu.
\newblock The role of the inhomogeneous demagnetizing field on the reversal
  mechanism in nanowire arrays.
\newblock {\em Journal of Physics D: Applied Physics}, 50(47):475002, 2017.

\bibitem{da2011reduction}
S~Da~Col, M~Darques, O~Fruchart, and L~Cagnon.
\newblock Reduction of magnetostatic interactions in self-organized arrays of
  nickel nanowires using atomic layer deposition.
\newblock {\em Applied Physics Letters}, 98(11):112501, 2011.

\bibitem{wang2008magnetic}
T~Wang, Y~Wang, Y~Fu, T~Hasegawa, H~Oshima, K~Itoh, K~Nishio, H~Masuda, FS~Li,
  H~Saito, et~al.
\newblock Magnetic behavior in an ordered co nanorod array.
\newblock {\em Nanotechnology}, 19(45):455703, 2008.

\bibitem{bochmann2018preparation}
S~Bochmann, D~D{\"o}hler, B~Trapp, M~Sta{\v{n}}o, O~Fruchart, and J~Bachmann.
\newblock Preparation and physical properties of soft magnetic nickel-cobalt
  three-segmented nanowires.
\newblock {\em Journal of Applied Physics}, 124(16):163907, 2018.

\bibitem{vivas2012magnetic}
LG~Vivas, J~Escrig, DG~Trabada, GA~Badini-Confalonieri, and M~V{\'a}zquez.
\newblock Magnetic anisotropy in ordered textured co nanowires.
\newblock {\em Applied Physics Letters}, 100(25):252405, 2012.

\bibitem{lavin2009angular}
R~Lavin, JC~Denardin, J~Escrig, D~Altbir, A~Cort{\'e}s, and H~G{\'o}mez.
\newblock Angular dependence of magnetic properties in ni nanowire arrays.
\newblock {\em Journal of Applied Physics}, 106(10):103903, 2009.

\bibitem{egolf2016hyperthermia}
Peter~W Egolf, Naveen Shamsudhin, Salvador Pan{\'e}, Didier Vuarnoz, Juho
  Pokki, Anne-Gabrielle Pawlowski, Paulin Tsague, Bastien de~Marco, William
  Bovy, Sinisa Tucev, et~al.
\newblock Hyperthermia with rotating magnetic nanowires inducing heat into
  tumor by fluid friction.
\newblock {\em Journal of Applied Physics}, 120(6):064304, 2016.

\bibitem{leulmi2015triggering}
Selma Leulmi, Xavier Chauchet, Melissa Morcrette, Guillermo Ortiz,
  H{\'e}l{\`e}ne Joisten, Philippe Sabon, Thierry Livache, Yanxia Hou, Marie
  Carri{\`e}re, St{\'e}phane Lequien, et~al.
\newblock Triggering the apoptosis of targeted human renal cancer cells by the
  vibration of anisotropic magnetic particles attached to the cell membrane.
\newblock {\em Nanoscale}, 7(38):15904--15914, 2015.

\bibitem{bonilla2017magnetic}
F~J Bonilla, L-M Lacroix, and T~Blon.
\newblock Magnetic ground states in nanocuboids of cubic magnetocrystalline
  anisotropy.
\newblock {\em Journal of Magnetism and Magnetic Materials}, 428:394--400,
  2017.

\bibitem{lillo2009pore}
M~Lillo and D~Losic.
\newblock Pore opening detection for controlled dissolution of barrier oxide
  layer and fabrication of nanoporous alumina with through-hole morphology.
\newblock {\em Journal of Membrane Science}, 327(1-2):11--17, 2009.

\bibitem{roslyakov2017growth}
I~V Roslyakov, D~S Koshkodaev, A~A Eliseev, D~Hermida-Merino, V~K Ivanov, A~V
  Petukhov, and K~S Napolskii.
\newblock Growth of porous anodic alumina on low-index surfaces of al single
  crystals.
\newblock {\em The Journal of Physical Chemistry C}, 121(49):27511--27520,
  2017.

\bibitem{rodriguez2001introduction}
Juan Rodr{\'\i}guez-Carvajal.
\newblock An introduction to the program fullprof 2000.
\newblock {\em Version July}, page~54, 2001.

\bibitem{zolotoyabko2009determination}
Emil Zolotoyabko.
\newblock Determination of the degree of preferred orientation within the
  march--dollase approach.
\newblock {\em Journal of applied Crystallography}, 42(3):513--518, 2009.

\bibitem{hertel2015analytic}
Riccardo Hertel and Attila K{\'a}kay.
\newblock Analytic form of transverse head-to-head domain walls in thin
  cylindrical wires.
\newblock {\em Journal of Magnetism and Magnetic Materials}, 379:45--49, 2015.

\bibitem{landeros2007reversal}
P~Landeros, S~Allende, J~Escrig, E~Salcedo, D~Altbir, and EE~Vogel.
\newblock Reversal modes in magnetic nanotubes.
\newblock {\em Applied Physics Letters}, 90(10):102501, 2007.

\bibitem{jamet2015head}
S~Jamet, N~Rougemaille, J-C Toussaint, and O~Fruchart.
\newblock Head-to-head domain walls in one-dimensional nanostructures: an
  extended phase diagram ranging from strips to cylindrical wires.
\newblock In {\em Magnetic Nano-and Microwires}, pages 783--811. Elsevier,
  2015.

\bibitem{fredkin1990hybrid}
DR~Fredkin and TR~Koehler.
\newblock Hybrid method for computing demagnetizing fields.
\newblock {\em IEEE Transactions on Magnetics}, 26(2):415--417, 1990.

\bibitem{fischbacher2007systematic}
T~Fischbacher, M~Franchin, G~Bordignon, and H~Fangohr.
\newblock A systematic approach to multiphysics extensions of
  finite-element-based micromagnetic simulations: Nmag.
\newblock {\em IEEE Transactions on Magnetics}, 43(6):2896--2898, 2007.

\bibitem{cullity2011introduction}
Bernard~Dennis Cullity and Chad~D Graham.
\newblock {\em Introduction to magnetic materials}.
\newblock John Wiley \& Sons, 2011.

\bibitem{ivanov2015micromagnetic}
Yurii~P Ivanov and O~Chubykalo-Fesenko.
\newblock Micromagnetic simulations of cylindrical magnetic nanowires.
\newblock In {\em Magnetic Nano-and Microwires}, pages 423--448. Elsevier,
  2015.

\bibitem{zighem2011dipolar}
F~Zighem, T~Maurer, F~Ott, and G~Chaboussant.
\newblock Dipolar interactions in arrays of ferromagnetic nanowires: A
  micromagnetic study.
\newblock {\em Journal of Applied Physics}, 109(1):013910, 2011.

\bibitem{panagiotopoulos2013packing}
I~Panagiotopoulos, W~Fang, F~Ott, F~Bou{\'e}, K~A{\"\i}t-Atmane, J-Y Piquemal,
  and G~Viau.
\newblock Packing fraction dependence of the coercivity and the energy product
  in nanowire based permanent magnets.
\newblock {\em Journal of Applied Physics}, 114(14):143902, 2013.

\bibitem{panagiotopoulos2015athermal}
I~Panagiotopoulos.
\newblock Athermal exploration of kagome artificial spin ice states by rotating
  field protocols.
\newblock {\em Journal of Magnetism and Magnetic Materials}, 384:70--74, 2015.

\bibitem{dobosz2019synthesis}
I~Dobosz, W~Gumowska, and M~Czapkiewicz.
\newblock Synthesis and magnetic properties of fe nanowire arrays
  electrodeposited in self-ordered alumina membrane.
\newblock {\em Archives of Metallurgy and Materials}, 64, 2019.

\end{thebibliography}

\end{document}